\documentclass[review,number,sort&compress]{elsarticle}

\usepackage{algorithm}
\usepackage{algorithmic}
\usepackage{amsmath} 
\usepackage{amssymb}
\usepackage{enumerate}
\usepackage{graphicx}  
\usepackage[ansinew]{inputenc} 
\usepackage{subfigure}
\usepackage[automake]{glossaries}
\makeglossaries
\newacronym{ac}{AC}{Air Channel}
\newacronym{dvfs}{DVFS}{Dynamic Voltage and Frequency Scaling}
\newacronym{fu}{FU}{Functional Unit}
\newacronym{ic}{IC}{Integrated Circuit}
\newacronym{lc}{LC}{Liquid Channels}
\newacronym{moea}{MOEA}{Multi-Objective Evolutionary Algorithm}
\newacronym{mfa}{MFA}{Multi-objective Floorplanning Algorithm}
\newacronym{pcb}{PCB}{Printed Circuit Board}
\newacronym{rc}{RC}{Resistance-Capacitance}
\newacronym{tsv}{TSV}{Through Silicon Via}
\usepackage{pifont}    
\usepackage{natbib}    
\usepackage{geometry}  
\usepackage{fleqn}     
\usepackage{txfonts}   
\usepackage{hyperref}  

\begin{document}

\title{Thermal-Aware Floorplanner for 3D IC, including TSVs, Liquid
  Microchannels and Thermal Domains Optimization}

\author[addr1]{David Cuesta} \ead{dcuestag@pdi.ucm.es}
\author[addr1]{José L. Risco-Martín\corref{cor1}} \ead{jlrisco@ucm.es}
\author[addr1]{José L. Ayala} \ead{jayala@ucm.es}
\author[addr1]{J. Ignacio Hidalgo} \ead{hidalgo@fis.ucm.es}
\cortext[cor1]{Corresponding author} \address[addr1]{Facultad de
  Informática, Universidad Complutense de Madrid (UCM), C/Prof. José
  García Santesmases s/n, 28040 Madrid, Spain}

\begin{abstract}
3D stacked technology has emerged as an effective mechanism to
overcome physical limits and communication delays found in 2D
integration. However, 3D technology also presents several drawbacks
that prevent its smooth application. Two of the major concerns are
heat reduction and power density distribution. In our work, we propose
a novel 3D thermal-aware floorplanner that includes: (1) an effective
thermal-aware process with 3 different evolutionary algorithms that
aim to solve the soft computing problem of optimizing the placement of
functional units and through silicon vias, as well as the smooth
inclusion of active cooling systems and new design strategies,(2)
an approximated thermal model inside the optimization loop, (3) an
optimizer for active cooling (liquid channels), and (4) a novel
technique based on air channel placement designed to isolate thermal
domains have been also proposed. The experimental work is conducted
for a realistic many-core single-chip architecture based on the
Niagara design. Results show promising improvements of the thermal and
reliability metrics, and also show optimal scaling capabilities to
target future-trend many-core systems.
\end{abstract}

\begin{keyword}
3D architecture \sep Thermal-aware floorplan \sep Air channels \sep
Through silicon vias \sep Evolutionary algorithms
\end{keyword}

\maketitle

\printglossaries

\section{Introduction}\label{sec:Introduction}
The process of continuous scaling over the past decades has led to
important improvements in size and performance of electronic products,
but it has also led to several challenges such as communication
problems and temperature management.

The shift to the many-core architectures has been driven by the
advances in semiconductor technologies. Besides, the increase in power
dissipation has been tailored by dynamic techniques like \gls{dvfs}
\cite{Choi2002}, using several clock domains \cite{Ogras2007} or task
migration policies \cite{Cuesta2010}. These techniques, however, may
also impact negatively the performance of the system. Design-time
approaches like the one proposed in this paper are able to mitigate
the effect of high temperatures and, also, are compatible with any
other existing dynamic technique applied to maintain or even increase
the performance of the system.

One important mechanism to overcome physical limits in \gls{ic} design
underlies on the design of multi-level \glspl{ic} using advanced
processes. These techniques boost the concept of three dimensional
integrated circuits (3D \glspl{ic}). 3D designs improve the
performance of the system by reducing interconnect delays and
increasing the density of the logic. \glspl{tsv} connect multiple
layers of the stack reducing distances between \glspl{fu}, hence
decreasing the communication delay. This fabrication technology also
allows the integration of multiple and disparate technologies, such as
radio frequency and mixed signal components, with traditional
computing technologies.

However, 3D integration exacerbates the problem of temperature in the
chip, specially temperature in inner layers. These thermal problems
are increasingly affecting the performance and the reliability of
electronic systems. \cite{Shang2004} reported that over 50\% of
electronic product failures are caused by thermal issues and the
presence of hotspots. Increasing the temperature decreases lifetime of
the chip exponentially. Furthermore, higher temperature can cause
slower devices, can increase leakage current, and can reduce the
performance due to the impact in the metal resistivity.

Considering these facts, it is desired to keep the components and the
chip structure as cool as possible for maximum reliability. However,
the absolute temperature of the chip is not the only factor that
affects performance; moreover, the thermal gradients that appear on
the chip surface degrade the system reliability through the promotion
of dangerous electro-migrations
\cite{Jonggook:TemperatureGradient:99}.

One way in which hardware designers have tried to address the thermal
problem is with the use of thermal-aware floorplanners such as
\cite{Cong2004} (that proposed a thermal-driven floorplanning
algorithm for 3D \glspl{ic}) or \cite{Healy2007} (where the authors
implemented a multi-objective floorplanning algorithm for 2D and 3D
\glspl{ic}, combining linear programming and simulated
annealing). Some other authors have also considered the placement of
thermal vias in these 3D stacks to optimize the thermal profile of
\glspl{ic} \cite{Wong2006}. The careful placement of active modules in
a 3D stack can optimize the wire length that connects \glspl{fu} by
reducing the communication delays, and can also provide a homogeneous
temperature distribution across the chip. Apart from static approaches
of thermal optimization, dynamic techniques are required to manage the
high power densities found in these architectures. While the
conventional air cooling has proved to be insufficient for
3D-\glspl{ic}, the interlayer micro-channel liquid cooling provides a
better option to address this problem. Some works like
\cite{DelValle2010} and \cite{Coskun2009} have focused in thermal
modeling with active cooling. These works have studied the effect of
allocating liquid channels between active layers and their cooling
effect.

\begin{figure*}[t]
\centering
\includegraphics[width=0.6\textwidth]{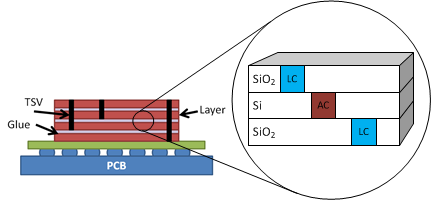}
\caption{3D architectural concept}
\label{fig:Concept}
\end{figure*}

Some of the goals in the design of 3D stacks are to achieve a
reduction in area and also to decrease the length of the
interconnections, that would be translated into improved data transfer
times and power consumption. Figure \ref{fig:Concept} summarizes our
proposed concept of a 3D-\gls{ic} architecture. Based on
\cite{Sabry2011}, the 3D stack is built over a \gls{pcb}, which is
considered to be adiabatic. Then, several layers are stacked, as can
be seen in Figure \ref{fig:Concept}. Every layer of the stack is
composed of silicon and silicon dioxide. \glspl{lc} is used as an
active cooling system. These channels are placed in the silicon
dioxide just over the active layers. Liquid channels absorb heat
produced by \glspl{fu} with a high power density, reducing the
temperature of inner layers. \glspl{lc} contain a coolant (generally
water) that is pumped into the chip. In this paper we also propose a
novel architectural set-up based on isolation channels or
\glspl{ac}. Air channels are etched in silicon and filled with low
pressure air. Since heat spread is mainly diffusive, air channels
prevent cold areas to be affected by the power dissipated in other
regions of the chip, creating temperature domains or regions. The use
of air channels has the major purpose of isolating thermal
regions. This could seem counter intuitive as the heat flow from
hotter modules could not be dissipated by the cooler ones. However,
the idea behind this is the optimization of the active cooling
mechanism (liquid channels), whose placement, number and required
cooling energy are benefited by the air channels. More in detail:

\begin{enumerate}
\item Since the chip is split in several independent regions, the
  thermal distribution of every domain can be defined attending to
  design constraints. In this way, it is easier to obtain a
  homogeneous distribution, or a pattern of alternate warm and cold
  regions, that help on achieving a more controlled thermal profile in
  the design phase.
\item If high temperatures are located in certain regions, dissipation
  mechanisms can be applied in those areas where the thermal problems
  are exacerbated, minimizing technological costs and optimizing the
  cooling properties of the different techniques.
\end{enumerate}

\begin{figure*}
\centering
\includegraphics[width=0.90\textwidth]{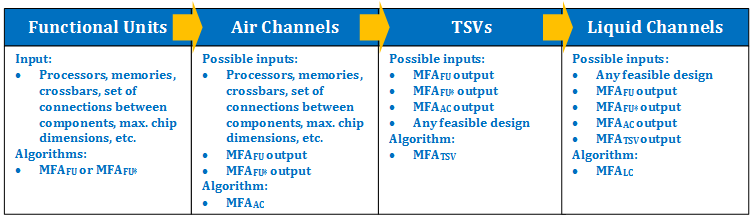}
\caption{3D design flow}
\label{fig:DesignFlow}
\end{figure*}

All this architectural diversity encourages the research on the design
of optimization algorithms to place automatically \glspl{fu},
\glspl{tsv}, as well as liquid and air channels, minimizing maximum
temperature and total wire length. The efficient management of all
these input variables, and the multiple optimization criteria given by
several objectives that have to be accomplished at the same time,
requires innovative algorithms capable of evaluating all these
parameters and returning the best set of solutions. In the case of 3D
\gls{ic} design, incremental optimization is a promising way to handle
multi-objective optimization with complicated constraints and to
facilitate the design reuse technology. Many works have been published
with this approach, however, none took into account thermal-aware
floorplanning. Most multi-objective floorplanning algorithms in the
literature are developed using Genetic Algorithms or Simulated
Annealing, where the main problem relies on the formulation of the
representation. Common representations for the floorplanning problem
are polish notation \cite{Berntsson2004}, combined bucket array
\cite{Cong2004} and O-tree \cite{Tang2007}. Most of these
representations do not perform well in our scenario because they were
initially developed to reduce area, whereas our problem is based on
minimizing temperature. To this end, we extend in this paper our
previous algorithm for floorplanning optimization named \gls{mfa}
\cite{Cuesta2013}, in which we formulated a multi-objective
optimization problem for 3D thermal aware floorplanning that is able
to reduce peak temperature, eliminate hotspots, and hence decrease
reliability risks related to temperature. \gls{mfa} allows the
incremental placement of \glspl{fu} and \glspl{tsv}.

In this paper, we will denote this algorithm with
\gls{mfa}$_{\mathrm{\gls{fu}}}$. From our previous work in
\cite{Cuesta2013}, as Figure \ref{fig:DesignFlow} introduces, we have
enhanced the \glspl{tsv} placement developing the new
\gls{mfa}$_{\mathrm{\gls{fu}*}}$ that slightly extends
\gls{mfa}$_{\mathrm{\gls{fu}}}$ to obtain solutions where at least one
\gls{tsv} configuration can be reached. We have also developed two new
evolutionary algorithms. These algorithms solve the integration of
active cooling systems within the 3D \gls{ic} using liquid channels by
optimizing their placement in those areas where the temperature is
higher. The second new algorithm can manage the optimization of
\gls{fu} and \gls{tsv} in the 3D \gls{ic} design with air channels,
taking into account new placement restrictions in the model. Results
show that the inclusion of isolation channels creates temperature
regions, decreasing the energy overhead imposed by the active cooling
system.

This paper makes major contributions in the area of thermal
optimization in 3D-integrated circuits. As compared with previous
approaches, the work presented here achieves a practical and effective
solution in the field of interest, outperforming the results obtained
by these works. This paper also extends our previous work presented in
\cite{Cuesta2013} with the following major upgrades:

\begin{itemize}
\item The proposal of a novel structure, called air isolation
  channels, as an effective mechanism for thermal isolation in 3D
  chips.
\item The development of the required thermal models for these
  structures that enable their control by the optimization algorithm.
\item The extension of the \gls{mfa} algorithm presented in
  \cite{Cuesta2013} with two new evolutionary algorithms, to consider
  the new design constraints and technologies, achieving better
  results in terms of thermal profile and fabrication costs.
\end{itemize}

The rest of the paper is structured as follows: Section
\ref{sec:ThermalModel} describes both the thermal model and current
implementation of \gls{mfa}. Then, Section \ref{sec:Mfa} continues
with an explanation of the proposed optimizer. The experimental set-up
used for our scenario is then described in Section \ref{sec:SetUp} and
finally, results and conclusions are presented in Sections
\ref{sec:Results} and \ref{sec:conclusions}, respectively.


\section{Thermal model}\label{sec:ThermalModel}

The equation governing heat diffusion via thermal conduction in a 3D
stack is \cite{Brooks2007}:

\begin{equation}\label{eq:HeatDiffusion}
\rho c \frac{\partial T\left( \vec{r}, t\right) }{\partial t} = \nabla \left( k(\vec{r}) \nabla T(\vec{r},t) \right) + p(\vec{r},t)
\end{equation}

subject to the boundary condition

\begin{equation}\label{eq:Boundaries}
k(\vec{r},t)\frac{\partial T(\vec{r},t)}{\partial n_i} + h_i T(\vec{r},t) = f_i (\vec{r},t)
\end{equation}

Regarding Equation \ref{eq:HeatDiffusion}, $\rho$ is the material
density, $c$ is the mass heat capacity, $T(\vec{r},t)$ and
$k(\vec{r})$ are the temperature and thermal conductivity of the
material at position $\vec{r}$ and time $t$, and $p(\vec{r},t)$ is the
power density of the heat source. With respect to Equation
\ref{eq:Boundaries}, $n_i$ is the outward direction normal to the
boundary surface $i$, $h_i$ is the heat transfer coefficient and $f_i$
is an arbitrary function at the surface $i$.

Numerical thermal analysis can be accomplished by applying a seven
points finite difference discretization method to Equation
\ref{eq:HeatDiffusion}, which decomposes the 3D stack into numerous
rectangular parallelepipeds of non-uniform sizes and shapes if
necessary. In this way, each element has a power dissipation,
temperature, thermal capacitance and thermal resistance to adjacent
elements, which interact via heat diffusion. The discretized equation
at an inner point of a grid element is:

\begin{eqnarray}
\rho c V \frac{T^{q+1}_{i,j,l}-T^{q}_{i,j,l}}{\Delta t} & = & -2(R_x+R_y+R_z)T^{q}_{i,j,l} + \nonumber \\
&	&	+ R_xT^{q}_{i-1,j,l} +\nonumber \\
                                                        &   & + R_xT^{q}_{i+1,j,l} + R_yT^{q}_{i,j-1,l} \nonumber \\
                                                        &   & + R_yT^{q}_{i,j+1,l} + R_zT^{q}_{i,j,l-1} \nonumber \\
                                                        &   & + R_zT^{q}_{i,j,l+1} + Vp_{i,j,l} \label{eq:DiscretizedHeat}
\end{eqnarray}

where $i$, $j$ and $l$ are discrete offsets along the $x$, $y$ and $z$
axes, $\Delta t$ is the discretization step in time $t$, $\Delta x$,
$\Delta y$ and $\Delta z$ are discretization steps along the $x$, $y$
and $z$ axes, and $V=\Delta x \Delta y \Delta z$. Finally, $R_x$,
$R_y$ and $R_z$ are the thermal conductivities between adjacent
elements, defined as follows:
$R_x=k\frac{\Delta_y\Delta_z}{\Delta_x}$,
$R_y=k\frac{\Delta_x\Delta_z}{\Delta_y}$, and
$R_z=k\frac{\Delta_x\Delta_y}{\Delta_z}$.

For a 3D stack with $N$ discretized elements, equation
\ref{eq:DiscretizedHeat} can be summarized as follows:

\begin{equation}\label{eq:ThermalTrans}
\mathbf{C}\frac{dT(t)}{dt} + \mathbf{R}T(t) = Pu(t)
\end{equation}

where the thermal capacitance matrix $\mathbf{C}$ is an $N \times N$
diagonal matrix, the thermal conductivity matrix $\mathbf{R}$ is an $N
\times N$ sparse matrix, $T(t)$ and $P$ are $N \times 1$ temperature
and power vectors, and $u(t)$ is the unit step function.

\begin{figure}[ht]
\centering
\subfigure[]{
    \includegraphics[width=0.45\textwidth]{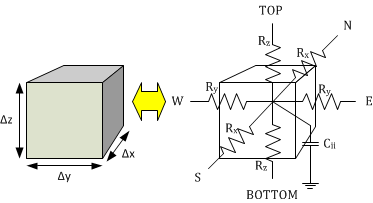}
    \label{fig:RegularCell}
}
\subfigure[]{
    \includegraphics[width=0.45\textwidth]{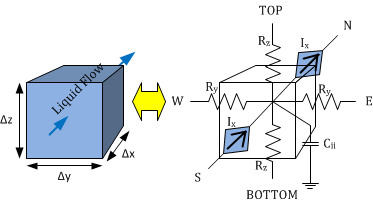}
    \label{fig:LiquidCell}
}
\caption[]{Material cells: \subref{fig:RegularCell} Diffusive cell
  (silicon, silicon dioxide, air) and \subref{fig:LiquidCell} Liquid
  cell.}
\label{fig:Cells}
\end{figure}

Equation \eqref{eq:ThermalTrans} can be characterized by a 3D \gls{rc}
model as the one presented in \cite{Ayala2012}. Voltage differences
are analogous to temperature differences, and the electrical
resistance is analogous to thermal resistance. The thermal modeling of
the stack in Figure \ref{fig:Concept} can be performed splitting the
chip into small cubic unitary cells.

Silicon and silicon dioxide cells are modeled with six thermal
resistances and one thermal capacitance as it can be seen in Figure
\ref{fig:RegularCell}. Four of these resistances connect each cell to
its lateral neighbors (those on the same layer), while the two
remaining resistances connect the cell with the upper and bottom cell,
respectively. The capacitance represents the heat storage inside the
cell.

Air channels also have a diffusive behavior. Despite the fact air is a
fluid, the dimensions of the cavity and the absence of forced
convection methods, make the fluid to behave as a diffusive
material. We have considered channels fabricated and filled with low
pressure air, which decreases thermal conductivity, making isolation
much more efficient.

Liquid channels are electrically modeled as it can be seen in Figure
\ref{fig:LiquidCell}. This difference with the diffusive cell comes
because, in cells circulating a coolant, the mechanism that dominates
the heat transfer is the forced convection. This mechanism can be
translated into our RC model using two voltage controlled current
sources, as it is profusely covered in \cite{Sridar2010}.

Heat diffusion to the surrounding environment is also considered by
the \gls{rc} model. This diffusion occurs in the edge of the chip and
in the top layer. Different chip packages and heat sinks can be
integrated in Equation \eqref{eq:ThermalTrans} by tuning the
capacitance and resistance parameters of the model. As the \gls{pcb}
base is considered to be adiabatic, no heat transfer occurs in the
bottom direction of the first layer.

The interface material that exists in between two silicon layers, used
as a glue, is modeled as an epoxy layer, a pure resistant
material. The existence of \glspl{tsv} is considered in the model,
also as a resistance element.

\begin{table*}
\centering
\begin{tabular}{ll}
\hline
Si linear thermal conductivity & 295 W/(mK) \\
Si quadratic term of thermal conductivity & -0.491 W/(m$K^2$) \\
SiO$_2$ thermal conductivity & 1.38 W/(mK) \\
Si specific heat & 1.628 x $10^6$ J/$m^3$K \\
SiO$_2$ specific heat & 4.180 x $10^6$ J/$m^3$K \\ 
Epoxy specific heat & 1.73 x $10^6$ J/$m^3$K\\
Epoxy thermal conductivity & 0.03 W/(mK)\\
\gls{tsv} thermal conductivity & 372 W/(mK)\\
\gls{tsv} specific heat & 3.45 x $10^6$ J/$m^3$K\\
Isolating Air thermal conductivity & 2.4 x $10^-3$ W/(mK)\\ 
Isolating Air specific heat & 1 x $10^4$ J/$m^3$K \\
Water specific heat & 4.184 x $10^6$ J/$m^3$K \\
Water thermal conductivity & 0.58 W/(mK) \\
\hline
\end{tabular}
\caption{Thermal properties of materials.}
\label{tab:Params}
\end{table*}

\begin{table}
\centering
\begin{tabular}{ll}
\hline
Stack width			&	12000 $\mu$m \\
Stack length	&	10500 $\mu$m \\
Cell size (lxw)	& 300x300 $\mu$m \\
\gls{tsv} cell size (lxw)	& 300x300 $\mu$m \\
Liquid channel width & 300 $\mu$m \\
SiO$_2$ height	& 50 $\mu$m \\
Si height	&	150  $\mu$m \\
Epoxy height	& 25 $\mu$m \\
\hline
\end{tabular}
\caption{Geometric properties of the stack.}
\label{tab:Dims}
\end{table}

All these cell types are integrated in Equation
\eqref{eq:ThermalTrans}, which is solved using an iterative method
(Forward Euler) to validate the results given by our 3D
floorplanner. The main thermal properties of the material and
dimensions used in the model are listed in Table \ref{tab:Params} and
\ref{tab:Dims}.

\section{Design Flow: Multi-objective Floorplanning Algorithm}\label{sec:Mfa}

\gls{mfa} was first proposed by David Cuesta \emph{et al.}
\cite{Cuesta2013}. This algorithm performs an incremental
floorplanning divided in two phases. First, \glspl{fu} are placed by
running \gls{mfa}$_{\mathrm{\gls{fu}}}$. Secondly,
\gls{mfa}$_{\mathrm{\gls{tsv}}}$ is executed, placing
\glspl{tsv}. These two processes are independent. Thus,
\gls{mfa}$_{\mathrm{\gls{fu}}}$ tends to obtain floorplans where the
insertion of \glspl{tsv} is not possible. In this section we describe
all the algorithms implied in the design flow shown in Figure
\ref{fig:DesignFlow}, including these two sub-algorithms for
self-content purposes. We also improve \gls{mfa}$_{\mathrm{\gls{fu}}}$
to reach feasible solutions for the insertion of \glspl{tsv}.


\subsection{\gls{mfa}$_{\mathrm{\gls{fu}}}$}

\gls{mfa}$_{\mathrm{\gls{fu}}}$ is a \gls{moea} based on NSGA-II
\cite{Deb2002}. The foorplanner manages coded solutions that are
gradually improved in the evolutionary process to provide
configurations optimized both in performance and thermal response for
the target architecture. To this end, \gls{mfa}$_{\mathrm{\gls{fu}}}$
simultaneously minimizes the following three objectives:
\begin{itemize}
\item $F_1$: Number of topological constraints violated (overlapping
  between different blocks, or components out of the borders of the
  chip).
\item $F_2$: Wire length, approximated as the Manhattan distance
  between interconnected blocks $\mathbf{C}$:
\begin{equation}\label{eq:F2}
F_2 = \sum_{i,j \in \mathbf{C} : i<j}{|x_i-x_j|+|y_i-y_j|+|z_i-z_j|}
\end{equation}
, where $(x_i,y_i,z_j)$ are the coordinates of \gls{fu} $i$.
\item $F_3$: Maximum temperature of the chip. The computation of this
  objective depends on the chosen thermal model. In the case of
  \gls{mfa}$_{\gls{fu}}$, the contribution to the maximum temperature
  of two \glspl{fu} $i,j$ is simplified as the cross product of their
  power densities $p_i, p_j$ divided by the corresponding euclidean
  distance. Thus, having $n$ \glspl{fu}, $F_3$ is defined as:

\begin{equation}\label{eq:F3}
F_3 = \sum_{i < j \in 1 \ldots n}{\frac{p_i \cdot
    p_j}{\sqrt{(x_i-x_j)^2+(y_i-y_j)^2+(z_i-z_j)^2}}}
\end{equation}

By minimizing $F_3$ the algorithm will try to place hottest blocks as
far as possible. This process reduces maximum temperature, as it is
demonstrated in \cite{Cuesta2013}.

\end{itemize}

\gls{mfa}$_{\mathrm{\gls{fu}}}$ can be classified as a hybrid
foorplanning approach because the decoding heuristic implements an
incremental foorplanning inspired in constructive techniques, while
the \gls{moea} on top of the heuristic is essentially iterative.

\begin{figure*}[ht]
\centering
\subfigure[]{
    \includegraphics[width=0.45\textwidth]{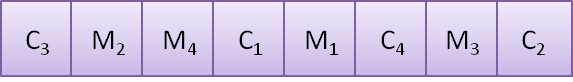}
    \label{fig:ChromosomeFU}
}
\subfigure[]{
    \includegraphics[width=0.617\textwidth]{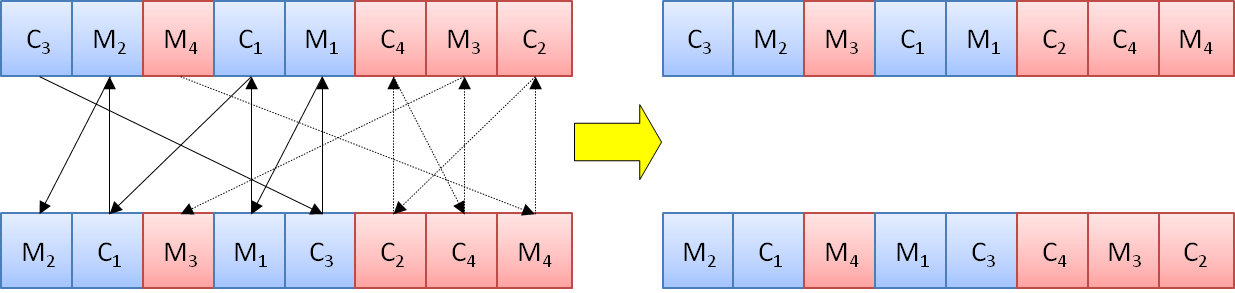}
    \label{fig:CrossoverFU}
}
\subfigure[]{
    \includegraphics[width=0.617\textwidth]{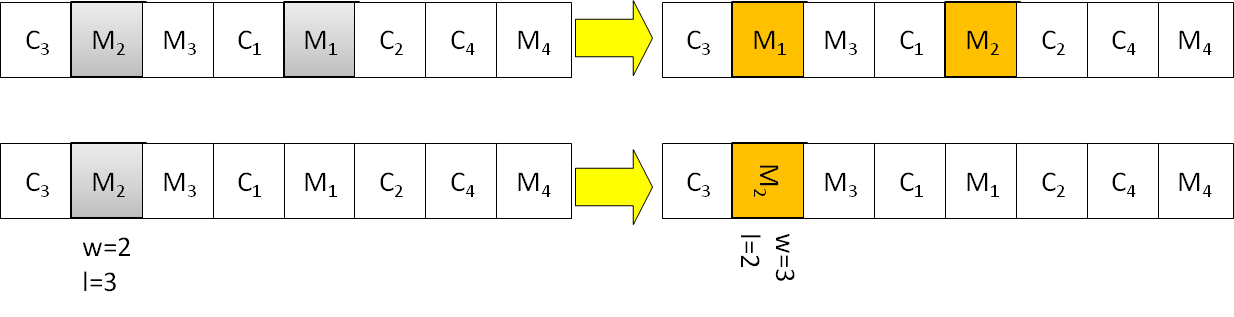}
    \label{fig:MutationFU}
}
\caption[]{\gls{mfa}$_{\mathrm{\gls{fu}}}$ A chromosome representing a
  solution of a platform with 4 cores, $C_i$, and 4 memories, $M_i$,
  \subref{fig:ChromosomeFU}, as well as crossover operation
  \subref{fig:CrossoverFU} and the two ways of mutation operator;
  swapping (up) and rotation of a FU (down) \subref{fig:MutationFU}
  operators.}
\label{fig:OperatorsFU}
\end{figure*}

We use a permutation encoding \cite{Sivanandam2007}, in which every
chromosome is a string of records representing the different
\glspl{fu} of the target architecture. These records gather
information relative to a \gls{fu}, namely label, width, and
length. Managing the width and length of the blocks allows to perform
rotations, granting further degrees of freedom to the optimization
process. Additional characteristics of the \glspl{fu} such as power
densities, connections, etc., must be managed by the
algorithm. However, this information does not need to be codified in
the chromosomes as it is common to all the individuals. Figure
\ref{fig:ChromosomeFU} depicts the representation of a chromosome used
in \gls{mfa}$_{\mathrm{\gls{fu}}}$. The example shows a candidate
solution (individual) of a platform composed of 8 \glspl{fu}: 4 cores
$C_i (i = 1, 2, 3, 4)$ and 4 memories $M_i (i = 1, 2, 3, 4)$. The
order $C_3,M_2,M_4,C_1,M_1,C_4,M_3,C_2$ determines the placement
sequence. Thus, $C_3$ will be placed first, followed by $M_2$, $M_4$
and so on.

All the chromosomes must have size $n$, where $n$ is the number of
\glspl{fu} to be placed. Therefore, the cardinality of the considered
solution space is $n!$. The operators designed according to the
representation are depicted in Figure \ref{fig:OperatorsFU} and
briefly described below:

\begin{itemize}
\item \emph{Selection:} The selection operator implements a binary
  tournament strategy. To this end, random couples of individuals are
  formed and the best solution of each pair is selected for crossover.

\item \emph{Crossover:} A cycle crossover is used to produce the
  offspring, this operator must take into account that all the
  components must appear once and only once in the chromosome (see
  Figure \ref{fig:CrossoverFU}).

\item \emph{Mutation:} The mutation of the solutions is performed in
  two ways. The first one consists in swapping the position of two
  blocks in the chromosome, resulting in a change of the placement
  sequence of the mutated individual. The effect of the second is the
  rotation of a \gls{fu} (see Figure \ref{fig:MutationFU}).
\end{itemize}

\begin{algorithm}[!t]
\caption{\gls{mfa}$_{\mathrm{\gls{fu}}}$}
\begin{algorithmic}
\begin{footnotesize}
\REQUIRE{$G$ is the number of generations. $N$ is the population size.}

\vspace{0.15cm}
\textbf{function} main()
\STATE $P$ = initialize() \COMMENT{$P$ is the first random population}
\STATE evaluate($P$) \COMMENT{$P$ is evaluated}
\FOR{$g=1$ \TO $G$}
  \STATE $\hat{P}=\emptyset$ \COMMENT{New empty population}
  \FOR{$n=1$ \TO $N/2$}
    \STATE $\hat{P_s}$ = select($P$) \COMMENT{Select two individuals,}
    \STATE $\hat{P_c}$ = crossover($\hat{P_s}$) \COMMENT{perform crossover ...}
    \STATE $\hat{P_m}$ = mutation($\hat{P_c}$) \COMMENT{and mutation}
    \STATE $\hat{P} = \hat{P} \cup \hat{P_m}$ 
  \ENDFOR
  \STATE evaluate($\hat{P}$)
  \STATE $P = P \cup \hat{P}$
  \STATE reduce($P$) \COMMENT{Standard NSGA-II reduction mechanism}
\ENDFOR

\vspace{0.15cm}
\textbf{function} evaluate($P$)
\FORALL{$I \in P$}
  \FOR{$i=1$ \TO $n$}
    \STATE $B_i \gets I_i$ \COMMENT{The i-th gene in individual $I$ $(I_i)$ represents the i-th block/functional unit $(B_i)$ to be placed in the current 3D candidate design}
    \STATE $f_i^* \gets \infty$, $x_i^* \gets 0, y_i^* \gets 0, z_i^* \gets 0$
    \\ \COMMENT{$0\leq l_i \leq L$, $0\leq w_i \leq W$ and $0\leq h_i \leq H$ are the length, width and height of block $i$, respectively}
    \FORALL{$(x_i \in [0..L-l_i],y_i \in [0..W-w_i],z_i \in [0..H-h_i])$}
      \STATE $F_1^i \gets $ checkTopologyConstraints($x_i,y_i,z_i,i$) \COMMENT{Number of topology constraints violated with the previous $i-1$ blocks already placed}
      \IF{$F_1^i = 0$}
        \STATE $F_2^i \gets$ manhattan($x_i,y_i,z_i,i$) \COMMENT{This function computes wire length according to Manhattan distances between connected blocks in the range $[1..i]$}
        \STATE $F_3^i \gets$ computeTemp($x_i,y_i,z_i,i$) \COMMENT{Compute \eqref{eq:F3} with $i<j\in 1..i$}
        \IF{$B_i$ is a core}
          \STATE $f_i \gets F_3^i$
        \ELSE
          \STATE $f_i \gets F_2^i$
        \ENDIF
        \IF{$f_i<f_i^*$}
          \STATE $f_i^* \gets f_i$
          \STATE $x_i^* \gets x_i, y_i^* \gets y_i, z_i^* \gets z_i$
        \ENDIF
      \ENDIF
    \ENDFOR
    \STATE $B_i \gets \left( x_i^*, y_i^*, z_i^* \right)$ \COMMENT{Assign best coordinates to each block}
  \ENDFOR
  \STATE $F_1 \gets $ checkTopologyConstraints() \COMMENT{Number of topology constraints violated in the current 3D \gls{ic}}
  \STATE $F_2 \gets$ manhattan() \COMMENT{Total wire length}
  \STATE $F_3 \gets$ computeTemp() \COMMENT{Compute \eqref{eq:F3}}  
  \STATE $I \gets \left( F_1, F_2, F_3 \right)$ \COMMENT{Assign multi-objective values to each individual}
\ENDFOR
\end{footnotesize}
\end{algorithmic}
\label{alg:MfaFu}
\end{algorithm}

Algorithm \ref{alg:MfaFu} shows the implementation of
\gls{mfa}$_{\gls{fu}}$. As Figure \ref{fig:DesignFlow} indicates, the
initial population is built from a list of components to be placed in
the floorplan, along with their dimensions and interconnectivity. Each
chromosome is defined just as a random sequence of these components
(see Figure \ref{fig:ChromosomeFU}). A heuristic is in charge of the
placement of the different elements of the architecture (decoding of
the solutions). The heuristic performs an incremental floorplanning in
which the components are sequentially placed in the 3D stack following
the order implied by the solution encoding. In fact, as the exhaustive
heuristic alone is capable of obtaining well performing solutions, the
\gls{moea} is designed to obtain the optimal order of the components
given the placement heuristic. In this heuristic, every block $i$ is
placed considering all the topological constraints, the wire length,
and the maximum temperature of the chip with respect to all the
previously placed blocks $j : j < i$, named as
$(F_1^i,F_2^i,F_3^i)$. The best location for each block is selected
depending on whether the block is a relative heat sink or a heat
source. For a heat source (like a core, for example) the best position
is the one with lowest $F_3^i$ to ensure an even thermal
distribution. If the block is a heat sink (like a memory) the best
position is the one with lowest wire length $F_2^i$. With this
procedure, the authors try to ensure a correct thermal
optimization. This approach is highly reasonable in terms of thermal
profile, since large 3D stacks with more than 48 cores reach
prohibitive temperatures (more than 400 K, as can be seen in
\cite{Cuesta2013}). Thus, every block is fixed in the remaining
position $(x_i, y_i, z_i)$. Once the placement has finalized, the
obtained configuration is evaluated according to the three defined
objectives $(F_1, F_2, F_3)$. In order to help the algorithm to find
feasible solutions, the multiobjective function can be transformed to
$(F_1, (1+F_1) \cdot F_2, (1+F_1) \cdot F_3)$. Although the first
objective can be removed in this case, we keep it to easly check if
the algorithm is not able to find feasible solutions.


\subsection{\gls{tsv} Optimization: \gls{mfa}$_{\mathrm{\gls{tsv}}}$}

As aforementioned, this algorithm is responsible of the placement of
\glspl{tsv} to allow vertical communications. As in
\gls{mfa}$_{\mathrm{\gls{fu}}}$, \gls{mfa}$_{\mathrm{\gls{tsv}}}$ is
based on NSGA-II. In the original version of \gls{mfa}, this algorithm
is the step executed immediately after
\gls{mfa}$_{\mathrm{\gls{fu}}}$. Since \glspl{tsv} insertion is not
checked in the first algorithm, it could happen that
\gls{mfa}$_{\mathrm{\gls{tsv}}}$ did not find a feasible distribution
of \glspl{tsv}. To solve this issue, we improve
\gls{mfa}$_{\mathrm{\gls{fu}}}$ in the next subsection. In the
following, we describe the \gls{mfa}$_{\mathrm{\gls{tsv}}}$ algorithm.

Technologically, and due to fabrication process constraints,
\glspl{tsv} can only be built from one specific layer to any other. In
this regard, \gls{mfa}$_{\mathrm{\gls{tsv}}}$ considers the top layer
as the initial one.

\begin{figure*}[ht]
\centering
\includegraphics[width=0.99\textwidth]{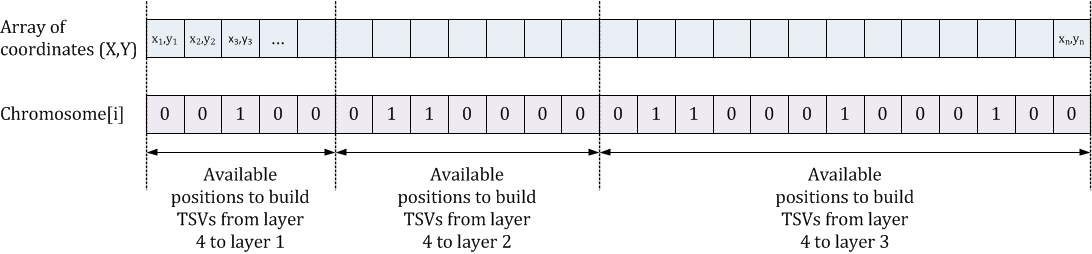}
\caption{\gls{tsv} Chromosome description.}
\label{fig:ChromosomeTSV}
\end{figure*}

To encode a solution, \gls{mfa}$_{\mathrm{\gls{tsv}}}$ examines the
remaining free cells in the \gls{mfa}$_{\mathrm{\gls{fu}}}$ resultant
stack. Then, \gls{mfa}$_{\mathrm{\gls{tsv}}}$ builds an array of x-y
coordinates where \glspl{tsv} can be drilled, as the array of
coordinates in Figure \ref{fig:ChromosomeTSV} shows. Given a 3D IC
with $N$ layers, a first region of this array contains the coordinates
where a \glspl{tsv} between layers $Top$ and $1$ can be built; a
second region in the same array contains the coordinates where
\glspl{tsv} between layers $Top$ and $2$ can be built, and so
forth. Next, a 0-1 chromosome of length equal to the array of
coordinates is created. If a gene contains a $1$, a \gls{tsv} is
inserted in the corresponding (x,y) position, between the layers that
belong to the corresponding region. In this way, Figure
\ref{fig:ChromosomeTSV} encodes $7$ \glspl{tsv} in four layers
($N=4$): 1 \gls{tsv} drilled between layers $4$ and $1$, $2$
\glspl{tsv} between layers $4$ and $2$, and $4$ \glspl{tsv} between
layers $4$ and $3$. As a result, the initial population is a set of
binary chromosomes randomly generated. The corresponding (x,y)
coordinates are stored in a separate array of coordinates.

Algorithm \ref{alg:MfaTsv} shows a pseudocode for
\gls{mfa}$_{\mathrm{\gls{tsv}}}$. This algorithm returns a set of
solutions, considering the number of \glspl{tsv} $F_4$ and the new
total wire length $F_5$. This set constitutes a Pareto front
approximation, and the designer will have the chance to select the
best solution in terms of economic cost and wire length reduction,
considering that a minimum number of \glspl{tsv} must be included in
the design in order to fulfill the communication constraints. The
minimum number of \glspl{tsv} is calculated attending to the
communication bandwidth needs of cores.  We have calculated the data
that is transferred by an FM modulation/demodulation application as
the one explained in \cite{Cuesta2010}. The minimum number of
\glspl{tsv} is given by the technological parameters of the
\glspl{tsv} and the volume of data to transfer \cite{Weldezion2009}.

\begin{algorithm}[!t]
\caption{\gls{mfa}$_{\mathrm{\gls{tsv}}}$}
\begin{algorithmic}
\begin{footnotesize}
\REQUIRE{$I$ is the current individual to be evaluated (see Figure \ref{fig:ChromosomeTSV}). $C$ is the set of connections in the floorplan.}

\vspace{0.15cm}
\textbf{function} evaluate($I$)
\STATE $F_4 \gets \sum_{i=1..N}I_i$ \COMMENT{Number of \glspl{tsv}.}
\STATE $F_5 \gets 0$
\FORALL{$(B_i,B_j) \in C$}
  \STATE $dz \gets |z_i - z_j|$
  \IF{$dz=0$}
    \STATE $d \gets $ manhattan($x_i,y_i,x_j,y_j$) \COMMENT{Manhattan distance between blocks $i$ and $j$}
    \STATE $F_5 \gets F_5 + d$ 
  \ELSE
    \STATE $d \gets $ findBestTsv($I,i,j$) \COMMENT{This functions takes all the candidate \glspl{tsv} in $I$ and compute the Manhattan distance in the path block $i$ $\rightarrow$ \gls{tsv} $\rightarrow$ block $j$. At the end, it returns the best distance}
  	\STATE $F_5 \gets F_5 + d$ 
  \ENDIF
\ENDFOR
\STATE $I \gets \left( F_4, F_5 \right)$ \COMMENT{Assign multi-objective values to this individual}

\end{footnotesize}
\end{algorithmic}
\label{alg:MfaTsv}
\end{algorithm}


\subsection{\gls{mfa}$_{\mathrm{\gls{fu}*}}$}
In this Section, we slightly modify \gls{mfa}$_{\mathrm{\gls{fu}}}$ to
obtain solutions where at least one \gls{tsv} configuration can be
reached by \gls{mfa}$_{\mathrm{\gls{tsv}}}$. To this end, we present
Algorithm \ref{alg:MfaFuTsv}. We only show the evaluation function
because the main function is identical to
\gls{mfa}$_\mathrm{\gls{fu}}$. The modification is performed including
the $R$ matrix. $R$ contains all the free cells in the 3D stack. Thus,
given a floorplan, it is quite easy to check if a \gls{tsv} can be
drilled to connect two blocks placed at different layers. If this is
not possible, $F_2^i$ is set to infinity.

\begin{algorithm}[!t]
\caption{\gls{mfa}$_{\mathrm{\gls{fu}*}}$}
\begin{algorithmic}
\begin{footnotesize}
\REQUIRE{$G$ is the number of generations. $N$ is the population size.}

\vspace{0.15cm}
\textbf{function} evaluate($P$)
\FORALL{$I \in P$}
  \STATE $R \gets 1$ \COMMENT{Matrix $L \times W \times H$ with free cells in the 3D-\gls{ic}, i.e., where \glspl{fu} can be placed}
  \FOR{$i=1$ \TO $n$}
    \STATE $B_i \gets I_i$ \COMMENT{The i-th gene in individual $I$ $(I_i)$ represents the i-th block $(B_i)$ to be placed in the current 3D candidate design}
    \STATE $f_i^* \gets \infty$, $x_i^* \gets 0, y_i^* \gets 0, z_i^* \gets 0$
    \FORALL{$(x_i \in [0..L-l_i],y_i \in [0..W-w_i],z_i \in [0..H-h_i])$}
      \STATE $F_1^i \gets $ checkTopologyConstraints($x_i,y_i,z_i,i$) \COMMENT{Number of topology constraints violated with the previous $i-1$ blocks already placed}
      \IF{$F_1^i = 0$}
        \STATE $F_2^i \gets$ manhattan($x_i,y_i,z_i,i,R$) \COMMENT{This function computes wire length according to Manhattan distances between connected blocks in the range $[1..i]$. It also asserts that a \gls{tsv} can be created if two blocks are placed on different layers (it is easily computed using $R$)}
        \STATE $F_3^i \gets$ computeTemp($x_i,y_i,z_i,i$) \COMMENT{Compute \eqref{eq:F3} with $i<j\in 1..i$}
        \IF{$B_i$ is a core}
          \STATE $f_i \gets F_3^i$
        \ELSE
          \STATE $f_i \gets F_2^i$
        \ENDIF
        \IF{$f_i<f_i^*$}
          \STATE $f_i^* \gets f_i$
          \STATE $x_i^* \gets x_i, y_i^* \gets y_i, z_i^* \gets z_i$
        \ENDIF
      \ENDIF
    \ENDFOR
    \STATE $B_i \gets \left( x_i^*, y_i^*, z_i^* \right)$ \COMMENT{Assign best coordinates to each block}
    \STATE update($R,B_i$)
  \ENDFOR
  \STATE $F_1 \gets $ checkTopologyConstraints() \COMMENT{Number of topology constraints violated in the current 3D \gls{ic}}
  \STATE $F_2 \gets$ manhattan() \COMMENT{Total wire length}
  \STATE $F_3 \gets$ computeTemp() \COMMENT{Compute \eqref{eq:F3}}  
  \STATE $I \gets \left( F_1, F_2, F_3 \right)$ \COMMENT{Assign multi-objective values to each individual}
\ENDFOR
\end{footnotesize}
\end{algorithmic}
\label{alg:MfaFuTsv}
\end{algorithm}

In the following subsections we present two new subalgorithms of
\gls{mfa}, named \gls{mfa}$_{\mathrm{\gls{lc}}}$ and
\gls{mfa}$_{\mathrm{\gls{ac}}}$. \gls{mfa}$_{\mathrm{\gls{lc}}}$ has
been developed to place liquid channels in the 3D
\gls{ic}. \gls{mfa}$_{\mathrm{\gls{ac}}}$ has been designed to divide
the 3D stack into regions isolated by air channels. It is worth noting
that the following two algorithms receive as input the resultant 3D
\gls{ic} obtained either by \gls{mfa}$_{\mathrm{\gls{fu}}}$ or
\gls{mfa}$_{\mathrm{\gls{fu}*}}$+\gls{mfa}$_{\mathrm{\gls{tsv}}}$, as
Figure \ref{fig:DesignFlow} shows.


\subsection{Liquid Channel Optimization: \gls{mfa}$_{\mathrm{\gls{lc}}}$}

Once the 3D \gls{ic} has been thermally optimized placing \glspl{fu}
or \glspl{tsv}, we run our accurate thermal model in Equation
\eqref{eq:ThermalTrans} to evaluate the temperatures in the chip,
computing the temperature matrix $T$. \glspl{tsv} have been already
incorporated into this equation as a set of cells disposed vertically
in the 3D \gls{ic}. The temperature values and the floorplan
information are the inputs to our proposed liquid channel optimizer,
shown in Algorithm \ref{alg:MfaLc}.

\begin{figure}[ht]
\centering
\includegraphics[width=0.75\textwidth]{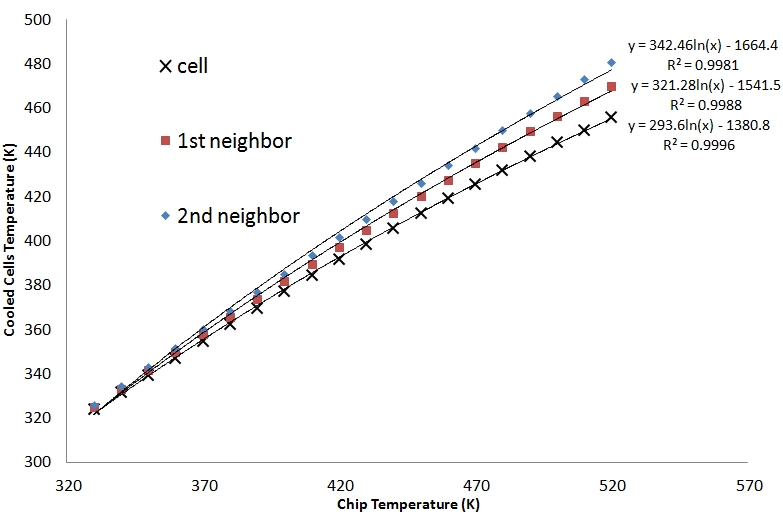}
\caption{Regression for evaluation function in liquid channels optimization}
\label{fig:Regression}
\end{figure}

Since liquid channels are placed right above \glspl{fu}, the only
topological constraint is the possible collision between liquid
channels and \glspl{tsv}. As was done for the placement of
\glspl{tsv}, a set of available $(x,z)$ coordinates is built. The
initial population is also a random set of binary chromosomes
indicating the presence or not of a liquid channel at a given
position. The evaluation function of \gls{mfa}$_{\mathrm{\gls{lc}}}$
takes into account the cooling effect of a liquid channel. In order to
evaluate this cooling effect, several thermal simulations were
conducted. Liquid channels are able to reduce temperature not only in
the active cells under the channel, but also in their neighbors. This
reduction follows a logarithmic tendency with the chip temperature
that can be seen in Figure \ref{fig:Regression}. Introducing these
data in the evaluation of our proposed evolutionary algorithm, we can
find optimized solutions.

The optimization of the number of microchannels in the design is a
major constraint from the point of view of technological and operating
costs. Adding liquid microchannels in the design implies not only
fabrication costs but also additional energy for the pumping system,
as shown in \cite{Changgu2011}.

\begin{algorithm}[!t]
\caption{\gls{mfa}$_{\mathrm{\gls{lc}}}$}
\begin{algorithmic}
\begin{footnotesize}
\REQUIRE{$I$ is the current individual to be evaluated. $T$ is the set
  of temperatures obtained with the thermal model.}

\vspace{0.15cm}
\textbf{function} evaluate($I$)
\STATE $F_6 \gets \sum_{i=1..N}I_i$ \COMMENT{Number of liquid channels.}
\STATE $F_7 \gets 0$
\STATE $\hat{T} \gets T$
\FORALL{$I_i \in I$}
  \IF{$I_i=1$}
    \STATE $(x_i, z_i) \gets I_i$ \COMMENT{Every gene is referred to a concrete $(x_i,z_i)$, where $z_i$ is the current layer for the i-th channel and $x_i$ its $x$ coordinate}
    \FOR{$y_i=0$ \TO $W-1$}
      \STATE $\hat{T}(x_i,y_i,z_i)=342.46ln(\hat{T}(x_i,y_i,z_i))-1664.4$
      \STATE $\hat{T}(x_i-1,y_i,z_i)=321.28ln(\hat{T}(x_i-1,y_i,z_i))-1541.5$
      \STATE $\hat{T}(x_i-2,y_i,z_i)=293.60ln(\hat{T}(x_i-2,y_i,z_i))-1380.8$
      \STATE $\hat{T}(x_i+1,y_i,z_i)=321.28ln(\hat{T}(x_i+1,y_i,z_i))-1541.5$
      \STATE $\hat{T}(x_i+2,y_i,z_i)=293.60ln(\hat{T}(x_i+2,y_i,z_i))-1380.8$
    \ENDFOR
  \ENDIF
\ENDFOR
\STATE $F_7 = \sum_{x_i,y_i,z_i}{\hat{T}((x_i,y_i,z_i))}$
\STATE $I \gets \left( F_6, F_7 \right)$ \COMMENT{Assign multi-objective values to this individual}
\end{footnotesize}
\end{algorithmic}
\label{alg:MfaLc}
\end{algorithm}


\subsection{Air Isolation Placement Optimization: \gls{mfa}$_{\mathrm{\gls{ac}}}$}

Our proposal of creating thermal domains in a chip is a revolutionary
method to keep heat concentrated in certain areas preventing, with air
isolation channels, heat spread from hot to cold regions. If this
design technique is combined with the deployment of liquid channels,
better results in terms of energy saving and fabrication cost are
obtained.

Since the thermal conductivity of air is lower than the silicon, the
channels can create thermal domains isolating regions with high
temperatures from other areas of the 3D stack. With this set up a more
homogeneous thermal distribution is obtained with a reduced investment
in any other active cooling mechanisms. The creation of thermal
domains implies a reduction in the number of liquid microchannels and
consequently a reduction in fabrication and working costs.

When the chip is isolated by air channels, the optimization placement
process is guided. This process works like the one described before,
but restricting certain areas of the chip to chosen \glspl{fu}
accordingly to their power consumption. Hot areas would be composed of
\glspl{fu} that have a high power density. On the other hand, elements
with a lower power consumption will be placed, all together, in cold
regions.

The considered algorithm is the same that the one described in
Algorithms \ref{alg:MfaFu} or \ref{alg:MfaFuTsv}, but different
topological constraints are included in the
\texttt{checkTopologyConstraints} function. Obviously, neither
\glspl{fu} nor \glspl{tsv} can occupy cells previously designed as air
channels. The initial population is generated as in
\gls{mfa}$_{\mathrm{\gls{fu}}}$ or
\gls{mfa}$_{\mathrm{\gls{fu}*}}$. However, as Figure
\ref{fig:DesignFlow} shows, feasible solutions obtained with these two
algorithms could be incorporated as a starting point for the
\gls{mfa}$_{\mathrm{\gls{ac}}}$ optimization process.

The definition of the topology is given by the designer and hence, is
an input to the optimization system as shown in Figure
\ref{fig:DesignFlow}. After that, it is the designer who specifies
where isolating air channels should be routed.


\section{Experimental set-up}\label{sec:SetUp}

Niagara2 and Niagara3 architectures are the base of our tests. This
distribution has been modified to include 48 SPARC cores. These 48
cores have been distributed in four layers, composed of 8-core
original Niagara2 in layers one and two and 16-core original Niagara3
in layers three and four. This scenario will be used in the following
for all the optimizations and simulations and also to compare
different optimization strategies.

The floorplan has also been modified in order to include an increased
number of cores which are placed in several layers of the 3D
stack. Since \gls{mfa} can place a variable number of cores in every
layer, the power consumption of the crossbar is scaled accordingly to
the number of cores found in every layer and their required
bandwidth. The inter-layer communication is resolved with a set of
\glspl{tsv} that route the communication signals from one layer to
another.

Worst case scenario has been set for power consumption. In our two
realistic floorplans, power consumption is set to 84W and 139W for
Niagara2 and Niagara3 respectively \cite{Niagara:Niagara}.

\begin{figure*}
\centering
\subfigure[]{
    \includegraphics[width=0.45\textwidth]{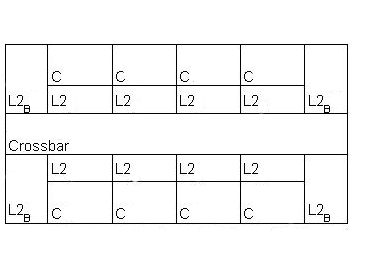}
    \label{fig:NiagaraCell}
}
\subfigure[]{
    \includegraphics[width=0.45\textwidth]{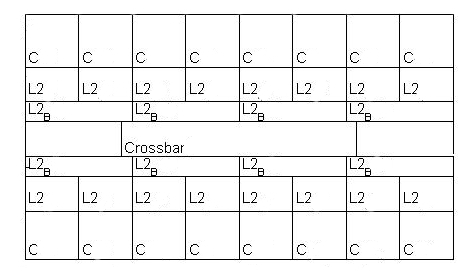}
    \label{fig:NiagaraaaCell}
}
\caption[]{Original floorplans \cite{Cuesta2013}}
\label{fig:originalflooplans}
\end{figure*}

As has been shown in section \ref{sec:Mfa},
\gls{mfa}$_\mathrm{\gls{fu}*}$ will place the \glspl{fu} that compose
the 3D multi-processor architecture to minimize the temperature
parameters. The area is set from the beginning of the optimization,
and the original distribution of components defines the area of the
optimization. The thermal results obtained by our floorplanner will be
compared with the stacks composed by the two original layers, based on
Niagara 2 and Niagara 3, presented in
Figure~\ref{fig:originalflooplans}. These two layers are disposed in
order to build a 48 core system. Niagara2 is placed in first and
second layer, and Niagara3 is used for layers 3 and 4. The cores (C),
memories (L2), shared memories (L2B) and crossbar (Crossbar) are
disposed in 4 layers.

Our experimental work will be focused on the analysis of the thermal
optimization achieved by the floorplanner and the additional
temperature reduction by the liquid cooling system. Additional
modifications have been conducted. We have created two isolated
thermal domains including air channels in the active layers. These
channels isolate an area of the chip creating ``hot islands'' that
will be then cooled with liquid channels. These air channels are
placed at 5400$\mu$m for layers 1 and 3 from the left side of the
chip, and the same distance for layers 2 and 4 from the right
side. Heat sources will be placed by the floorplanner in these areas,
whereas \glspl{fu} with lower power density will be placed in the warm
region.

On the other hand, \glspl{mfa} are configured with different
parameters. Both \gls{mfa}$_\mathrm{\gls{fu}*}$ and
\gls{mfa}$_\mathrm{\gls{ac}}$ are configured with a population of one
hundred individuals and number of generations equal to the number of
\glspl{fu}, which prevents the algorithm from stacking in a local
optimum. In both algorithms, crossover probability is set to 0.90 and
the mutation probability is set to 1/number of \glspl{fu} as
recommended in \cite{Deb2002}.

\gls{mfa}$_\mathrm{\gls{tsv}}$ and \gls{mfa}$_\mathrm{\gls{lc}}$ are
configured with a maximum population of one hundred individuals, and a
maximum number of 250 generations. The probability of mutation is set
depending on the number of variables; in this particular case, it is
the inverse of the number of available points to insert \glspl{tsv} or
liquid channels, respectively. Then, we set a single point crossover
with a probability of 0.90 and the tournament selection method,
following the guidelines given in \cite{Deb2002}.

\section{Results}\label{sec:Results}
This section presents the thermal results obtained in the scenarios
described in Section \ref{sec:SetUp}. All the thermal values have been
calculated using the thermal model described in section
\ref{sec:ThermalModel}. All the designs presented in this section have
been computed with \gls{mfa}$_{\mathrm{\gls{fu}*}}$,
\gls{mfa}$_{\mathrm{\gls{ac}}}$, \gls{mfa}$_{\mathrm{\gls{tsv}}}$ and
\gls{mfa}$_{\mathrm{\gls{lc}}}$.

In addition to maximum temperature (in $K$) and total wirelength (in
cells, each cell has 300 $\mu$m), we also include thermal gradient as
a measure of reliability. Maximum temperature and wire length are
presented as absolute values, thermal gradient is averaged among the
layers.

\begin{figure*}  
\centering
\includegraphics[width=0.95\textwidth]{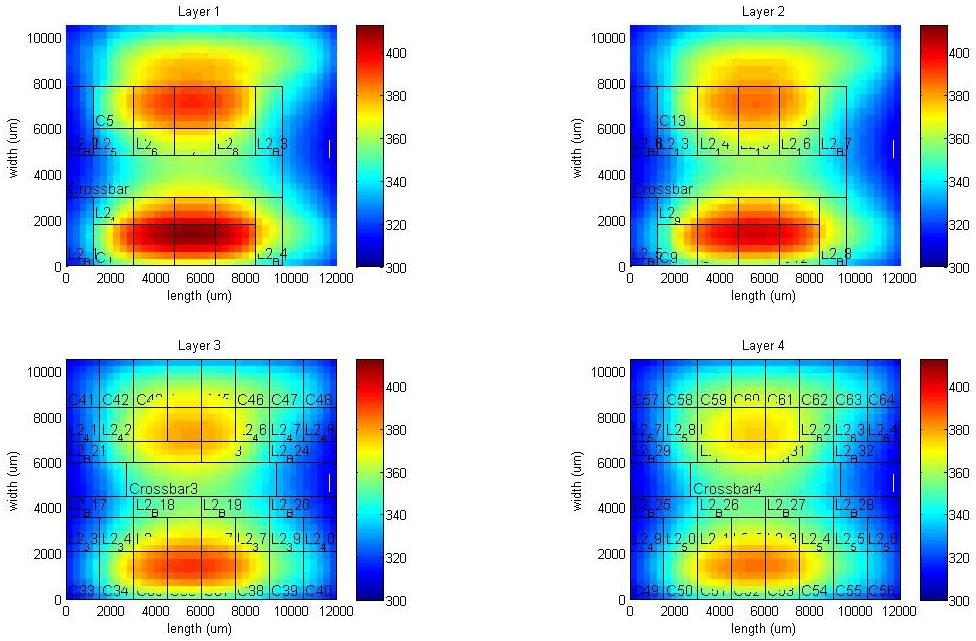}
\caption{Thermal maps of the original 48-core system \cite{Cuesta2013}}
\label{fig:original}
\end{figure*}

Figure \ref{fig:original} depicts the thermal distribution of our
baseline scenario, where all the cores are labeled as C$\{\mathrm{id}\}$,
memories as L2$_{\{\mathrm{id}\}}$, shared memories as L2$_{\mathrm{B}}$~
$\{\mathrm{id}\}$, and Crossbars as Crossbar$\{\mathrm{id}\}$, where \emph{id} is an identifier. As can be seen in the
figure, hot areas appear, specially in first and second layers because
they cannot dissipate heat as easy as top layers do.

\subsection{Functional units and \gls{tsv} placement: \gls{mfa}$_{\mathrm{\gls{fu}*}}$ + \gls{mfa}$_{\mathrm{\gls{tsv}}}$}

\begin{figure}
\centering
\subfigure[]{
    \includegraphics[width=0.45\textwidth]{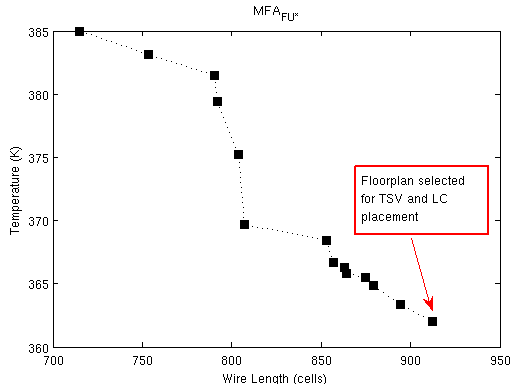}
    \label{fig:MfaFuStar}
}
\subfigure[]{
    \includegraphics[width=0.45\textwidth]{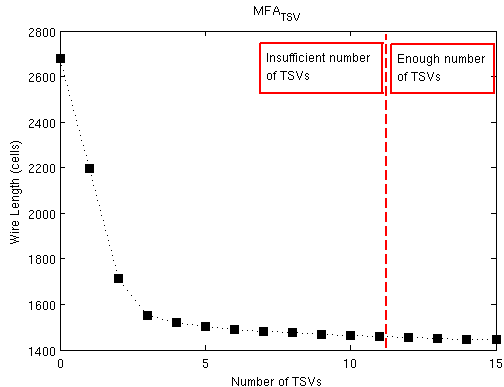}
    \label{fig:MfaTsv}
}
\caption[]{\subref{fig:MfaFuStar} Set of non-dominated solutions
  obtained with \gls{mfa}$_{\mathrm{\gls{fu}*}}$ and the floorplan
  selected to be used in \gls{mfa}$_{\mathrm{\gls{tsv}}}$ and
  \gls{mfa}$_{\mathrm{\gls{tsv}}}$. \subref{fig:MfaTsv} Set of
  non-dominated solutions reached by \gls{mfa}$_{\mathrm{\gls{tsv}}}$
  with the previous selected floorplan}
\label{fig:MfaFuStarAndTsv}
\end{figure}

Figure \ref{fig:MfaFuStarAndTsv} shows the non-dominated solutions
obtained by both \gls{mfa}$_{\mathrm{\gls{fu}*}}$ and
\gls{mfa}$_{\mathrm{\gls{tsv}}}$. Figure \ref{fig:MfaFuStar} shows the
resultant non-dominated front obtained by
\gls{mfa}$_{\mathrm{\gls{fu}*}}$. It is worth noting that the wire
length is not realistic in the sense than the Manhattan distance has
been also applied vertically, without adding \glspl{tsv} yet. One of
these non-dominated solutions must be selected by the system designer
in order to apply both \gls{mfa}$_{\mathrm{\gls{tsv}}}$ and
\gls{mfa}$_{\mathrm{\gls{lc}}}$ algorithms. In our case, we have
selected the one with the lowest temperature, as illustrated in Figure
\ref{fig:MfaFuStar}. Figure \ref{fig:MfaTsv} shows the non-dominated
solutions obtained by \gls{mfa}$_{\mathrm{\gls{tsv}}}$ with the number
of TSVs and chip wire length. The designer whill have to choose which
solution is more convenient in every case, depending on the economic
cost and technological issues. In out case, we have selected the point
$(11, 1459)$, since it fulfills the bandwith requirements for this
particular scenario \cite{Cuesta2013}.

\begin{figure*}  
\centering{\includegraphics[width=0.95\textwidth]{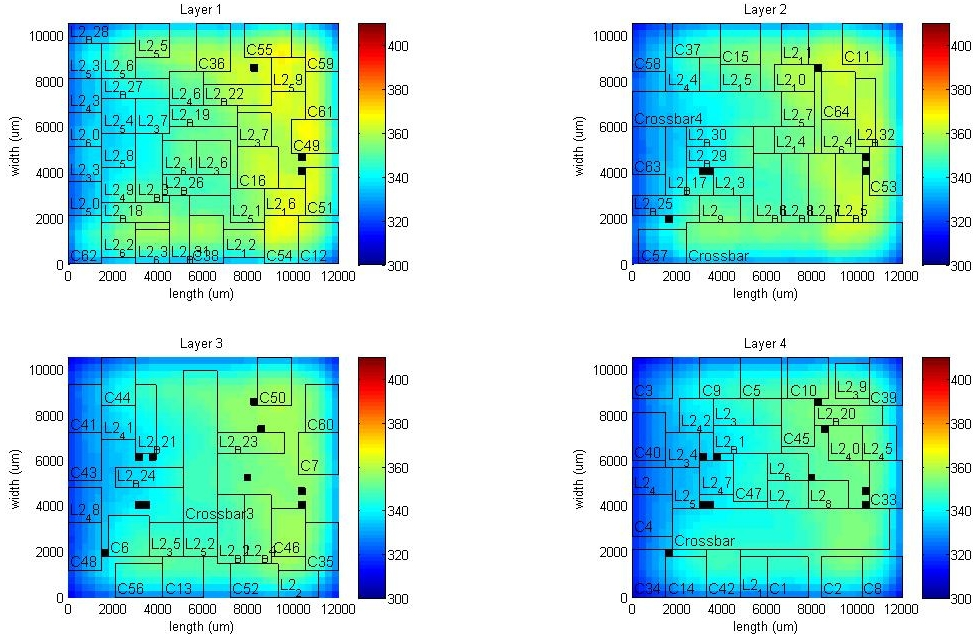}}
\caption{Thermal maps of the optimized 48-core system. Black spots
  show the position of the TSVs.}
\label{fig:48cores_optimizer}
\end{figure*}

Finally, Figure \ref{fig:48cores_optimizer} depicts the thermal maps
of the solution selected in (\gls{mfa}$_{\mathrm{\gls{fu}*}}$ and
\gls{mfa}$_{\mathrm{\gls{tsv}}}$). Black spots in the figures show the
position of the \glspl{tsv}. As can be seen in the Table
\ref{tab:comparison}, we achieve a reduction of 38 K in maximum
temperature when compared with the original scenario.

Table \ref{tab:comparison} compares the metrics obtained by the
original 3D floorplan with the one that has been optimized with
\gls{mfa}$_{\mathrm{\gls{fu}*}}$+\gls{mfa}$_{\mathrm{\gls{tsv}}}$. It
is worth noting that the wiring of both the Niagara2 and Niagara3
architectures is isolated by layer in our baseline stack (no vertical
communication occurs), highly optimized by the 2d floorplanner, and
thus impossible to improve. For this reason, the total wire length has
not been improved by our proposal. In the same manner, the wire length
computed by \gls{mfa}$_{\mathrm{\gls{fu}*}}$ (912 cells) is not
included in Table \ref{tab:comparison} because it is not realistic,
i.e., the Manhattan distance cannot be applied between different
layers in a realistic design.

\begin{table*}
\centering
\begin{tabular}{cccc}
\hline
\textbf{Algorithm}   & \textbf{Max.Temp. (K)} & \textbf{Grad. (K)} & \textbf{Wirelength (Cells)}\\
\hline
Baseline Stack ($\emptyset$)    & 399     & 89                  & 1012                 \\
\gls{mfa}$_{\mathrm{\gls{fu}*}}$            & 362     & 43                  & -                    \\ 
$\ldots$ + \gls{mfa}$_{\mathrm{\gls{tsv}}}$ & 362     &	43                  & 1459                 \\
\hline
\end{tabular}
\caption{Thermal and wire length comparison for the placement of \glspl{fu} and \glspl{tsv}}
\label{tab:comparison}
\end{table*}

As can be seen, our proposal is able to reduce chip temperature up to
37 K. However, as Table \ref{tab:comparison} shows, since we are
connecting \glspl{fu} in different layers the performance is worse
than in the baseline 3D stack, even when including
\glspl{tsv}. However, if we compute the total wirelength using wire
bonding as the inter-layer communication mechanism we could reach up
to 2680 cells in wire length. This proves the convenience of using
\glspl{tsv} as the most promising solution to take advantage of 3D
integration technology.

\subsection{Liquid microchannel optimization: \gls{mfa}$_{\mathrm{\gls{lc}}}$}

\begin{figure}
\centering
\subfigure[]{
    \includegraphics[width=0.45\textwidth]{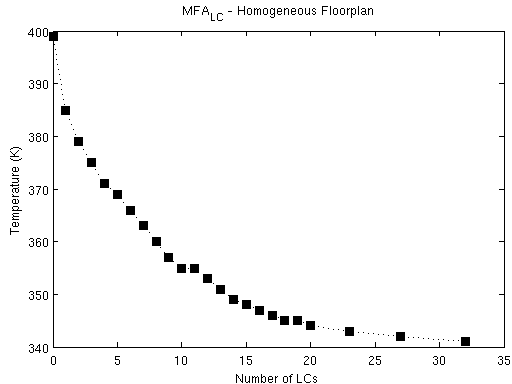}
    \label{fig:MfaLcHom}
}
\subfigure[]{
    \includegraphics[width=0.45\textwidth]{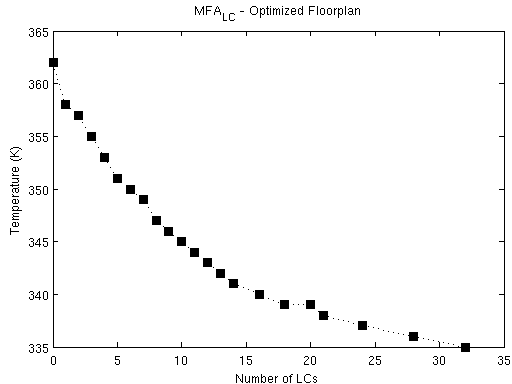}
    \label{fig:MfaLcOpt}
}
\caption[]{\subref{fig:MfaLcHom} Set of non-dominated solutions obtained with \gls{mfa}$_{\mathrm{\gls{lc}}}$ over the baseline stack. \subref{fig:MfaLcOpt} Set of non-dominated solutions reached by \gls{mfa}$_{\mathrm{\gls{lc}}}$ over the floorplan selected in \gls{mfa}$_{\mathrm{\gls{fu}*}}$}
\label{fig:MfaLc}
\end{figure}

The high temperatures reached in the baseline stack can be managed
including liquid micro channels, as an effective way to decrease
temperature in inner layers. In this scenario, 32 liquid channels were
placed. To this end, we executed \gls{mfa}$_{\mathrm{\gls{lc}}}$ with
$F_6 \leq 32$, obtaining the set of non-dominated solutions depicted
in Figure \ref{fig:MfaLc}. \gls{mfa}$_{\mathrm{\gls{lc}}}$ placed
liquid channels in both the baseline stack and the optimized floorplan
given by \gls{mfa}$_{\mathrm{\gls{fu}*}}$ +
\gls{mfa}$_{\mathrm{\gls{tsv}}}$. Obviously, the configuration with
more liquid channels obtained the lowest temperature.

\begin{figure*}  
\centering
\includegraphics[width=0.95\textwidth]{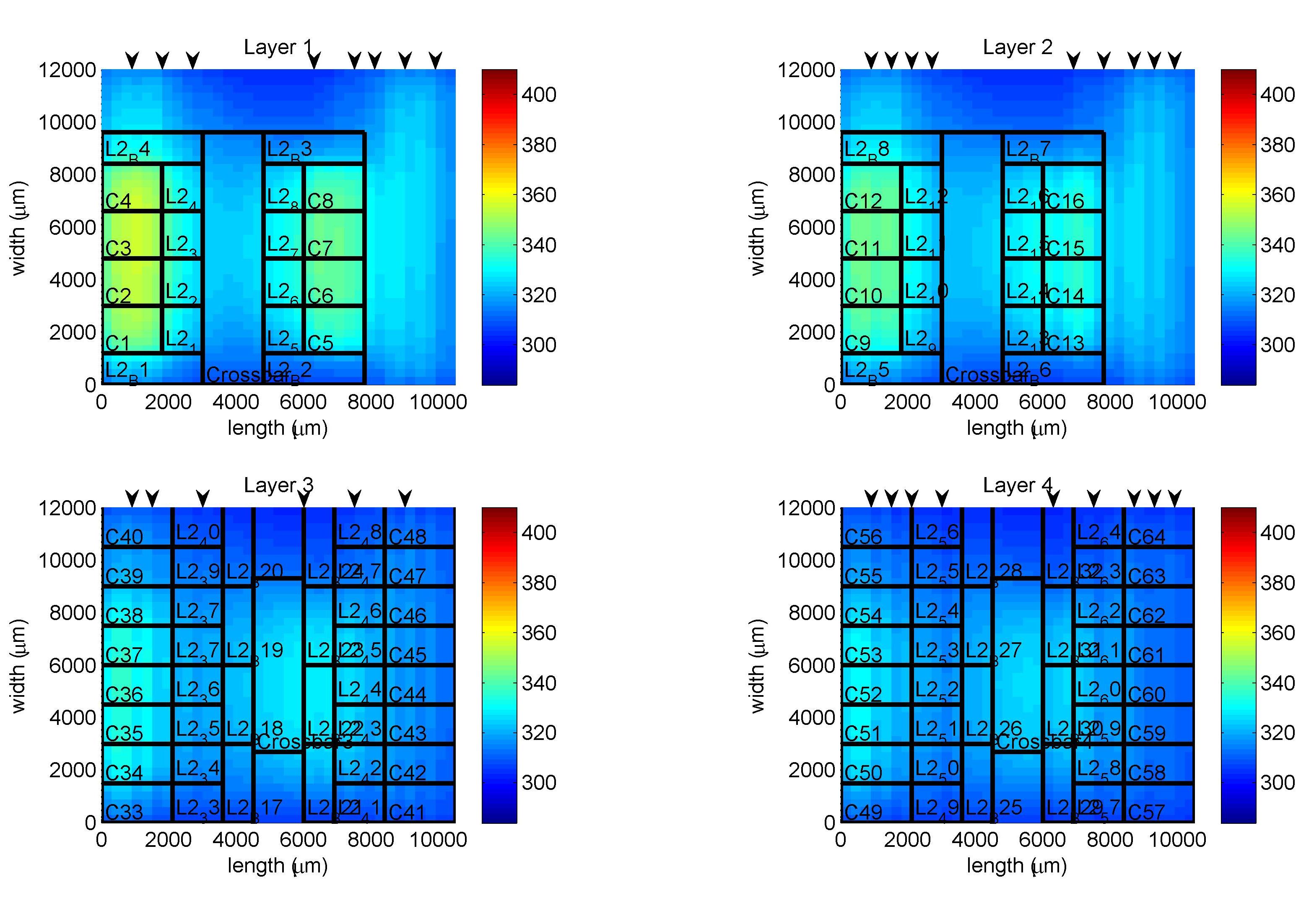}
\caption{Thermal maps of the original system with 32 optimized liquid channels. Arrowheads show the location of each channel.}
\label{fig:original_32optimized}
\end{figure*}

The optimized liquid channel placement thermogram can be seen in
Figure \ref{fig:original_32optimized} for the baseline stack. As can
be seen in the Figure, hot areas have disappeared because channels can
cool down the heat produced by the cores. The comparison results of
placing channels following a homogeneous distribution (8
channels/layer), and optimizing their position using
\gls{mfa}$_{\mathrm{\gls{lc}}}$ can be seen in table
\ref{tab:original_channels}. The homogeneous distribution of liquid
channels is able to reduce the temperature throughout the chip,
however, optimizing channel placement improves maximum temperature and
gradients.

\begin{table*}
\centering
\begin{tabular}{cccc}
\hline
\textbf{\gls{lc} Algorithm} & \textbf{Max.Temp. (K)} & \textbf{Grad. (K)} & \textbf{Wirelength (Cells)} \\
\hline
Baseline Stack ($\emptyset$) & 399 & 89 & 1012 \\
Homogeneous ($\emptyset$)    & 348 & 46 &	1012 \\
\gls{mfa}$_{\mathrm{\gls{lc}}}$          & 341 & 38 & 1012 \\
\hline
\end{tabular}
\caption{Homogeneous and optimized liquid channels in the original distribution.}
\label{tab:original_channels}
\end{table*}

Finding an optimal distribution for liquid channels improves not only
thermal metrics, but also fabrication costs, because wider channels
can be built to replace two or more thinner ones, also reducing
pumping energy. The optimization of the placement of liquid channels
achieves very good results, but these thermal results can be enhanced
adding several optimizations such as \glspl{fu} placement.

\begin{figure*}  
\centering
\includegraphics[width=0.95\textwidth]{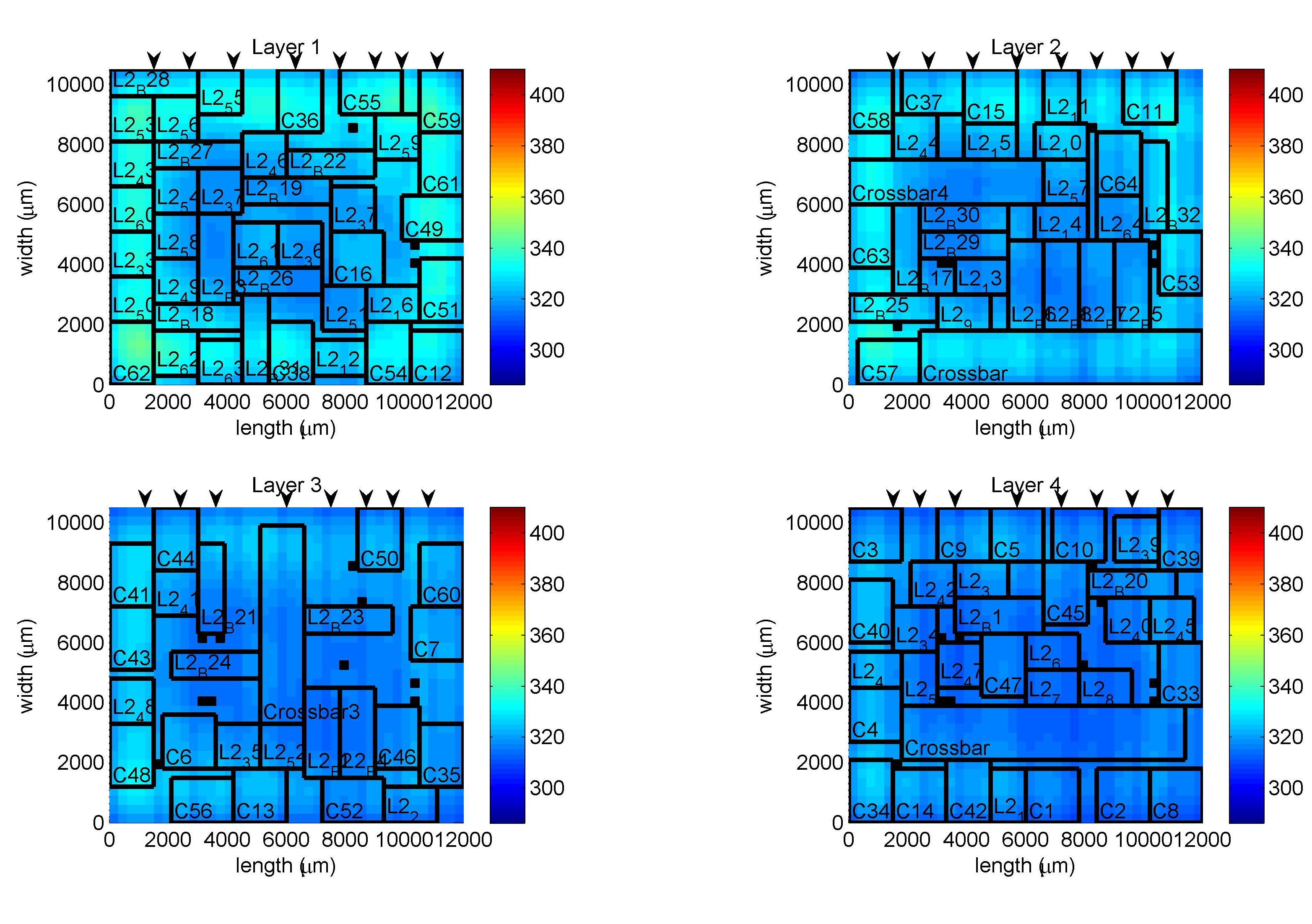}
\caption{Thermal maps of the optimized system with 32 optimized liquid channels. Black spots show the position of the TSVs. Arrowheads show the location of each channel.}
\label{fig:48cores_optimizer_32channels}
\end{figure*}

Let us now examine the placement of liquid channels over the optimized
floorplan. As was previously commented, liquid placement optimizer
evaluates not only hot areas but also \glspl{tsv} location, because
liquid channels cannot go through \glspl{tsv}. In Figure
\ref{fig:48cores_optimizer_32channels} it can be seen that once the
liquid channels are deployed, there are no longer hot
areas. Optimizing both the placement of \glspl{fu}, and the placement
of liquid microchannels, we have achieved a reduction of 57 K in
maximum temperature, 33 K in mean temperature and 66 K in the gradient
when compared with the original scenario.

Table \ref{tab:optimized_channels} shows data for the optimized
placement of \glspl{fu}, and the effect of adding 32 liquid channels
following a homogeneous and optimized placement. Placing channels in
an optimal way improves now every thermal metric. Including liquid
channels does not change the wirelength since the routing is
invariant.

\begin{table*}
\centering
\begin{tabular}{cccc} 
\hline
\textbf{\gls{lc} Algorithm} & \textbf{Max.Temp. (K)} & \textbf{Grad. (K)} & \textbf{Wirelength (Cells)} \\
\hline
\gls{mfa}$_{\mathrm{\gls{fu}*}}$ + $\emptyset$             & 362 & 43 & 1459 \\
\gls{mfa}$_{\mathrm{\gls{fu}*}}$ + homogeneous placement & 339 & 26 & 1459 \\
\gls{mfa}$_{\mathrm{\gls{fu}*}}$ + \gls{mfa}$_{\mathrm{\gls{lc}}}$   & 335 & 24 & 1459 \\
\hline
\end{tabular}
\caption{Homogeneous and optimized 32 liquid channels in the optimized distribution.}
\label{tab:optimized_channels}
\end{table*}

\subsection{Air channel isolation: \gls{mfa}$_\mathrm{\gls{ac}}$}

One of the main contributions of this work is adding air channels to
isolate thermal domains in the chip. This isolation makes the cooling
process much easier, because liquid channels can be placed in those
areas where the temperature is higher, reducing the number of deployed
liquid microchannels to achieve the same thermal profile. This
reduction in the number of microchannels implies a reduction in
fabrication costs and in the energy designated for the pumping system.

\begin{figure*}[ht]
\centering
\includegraphics[width=0.95\textwidth]{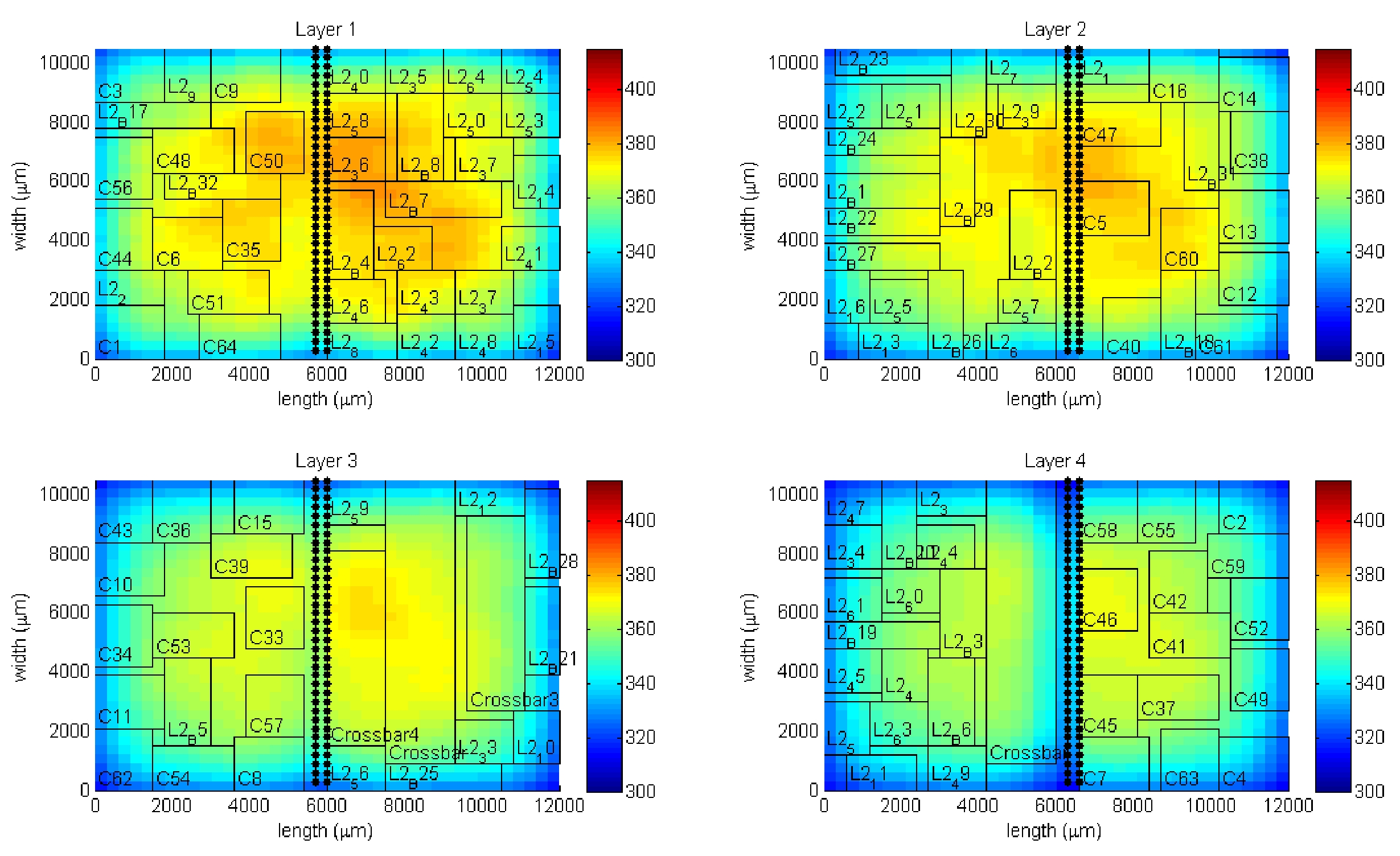}
\caption{Floorplan distribution with air isolation (\gls{mfa}$_{\mathrm{\gls{ac}}}$). $*$ symbols limit the location of the air channel.}
\label{fig:48cores_air}
\end{figure*}

\begin{figure*}[ht]
\includegraphics[width=0.95\textwidth]{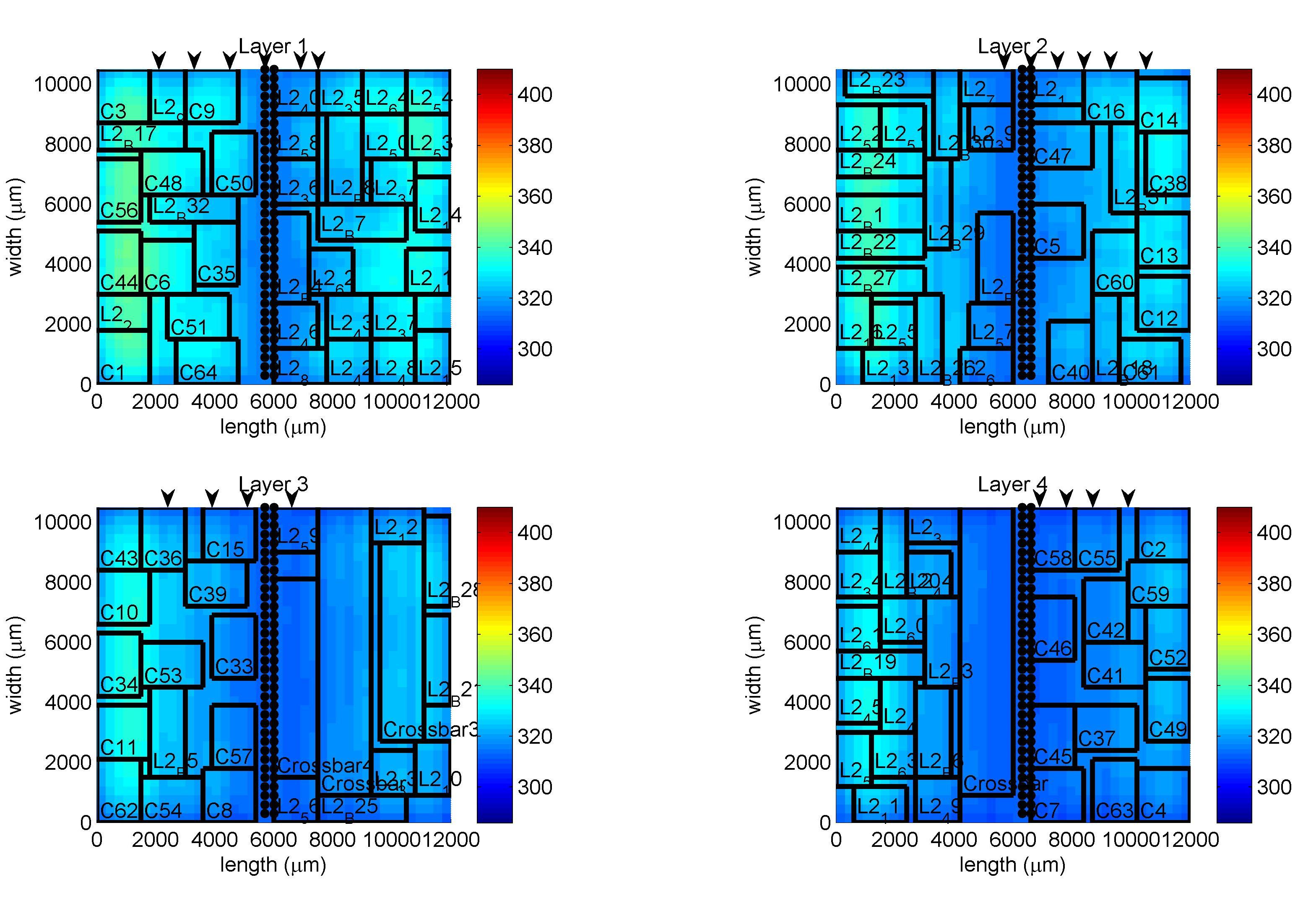}
\caption{Resultant floorplan with air isolation and 20 liquid channels (\gls{mfa}$_{\mathrm{\gls{ac}}}$ + \gls{mfa}$_{\mathrm{\gls{lc}}}$) $*$ symbols limit the location of the air channel.  Arrowheads show the position of liquid channels.}
\label{fig:48cores_air_20channel}
\end{figure*}

As was explained in section \ref{sec:SetUp} a hot region and a warm
region are created. Figure \ref{fig:48cores_air} presents the
floorplan in the non-dominated front obtained by
\gls{mfa}$_{\mathrm{\gls{ac}}}$ with the lowest temperature and
without liquid channels, where $*$ symbols represent the existence of
an air channel. In this regard, Figure
\ref{fig:48cores_air_20channel}, shows the same scenario but after
inserting 20 liquid channels with \gls{mfa}$_{\mathrm{\gls{lc}}}$. As
can be seen in Figure \ref{fig:48cores_air}, the floorplanner has
located \glspl{fu} with a high power consumption in hot regions (left
side for layers 1 and 3; right side for layers 2 and 4) and \glspl{fu}
with lower power densities in warm regions. Once the placement has
been done, liquid channels can be deployed in the 3D \gls{ic} (see
Figure \ref{fig:48cores_air_20channel}). Most of the optimized
channels are placed in hot regions but some of them are located with
heat sinks units in order to obtain a homogeneous thermal
distribution.

\begin{table*}
\centering
\begin{tabular}{ccccc}
\hline
\textbf{Floorplanner} & \textbf{\gls{lc} placement} & \textbf{\# of channels} & \textbf{Max.Temp. (K)} & \textbf{Grad. (K)} \\
\hline
\gls{mfa}$_{\mathrm{\gls{ac}}}$  & -                   &	0 &	377 & 61 \\
\gls{mfa}$_{\mathrm{\gls{fu}*}}$ & Homogeneous         & 32 & 335 & 24 \\
\gls{mfa}$_{\mathrm{\gls{ac}}}$  & \gls{mfa}$_{\mathrm{\gls{lc}}}$ & 20 & 335 &	27 \\
\hline
\end{tabular}
\caption{Comparison between optimized floorplan with and without air channel isolation plus liquid microchannels (\gls{mfa}$_{\mathrm{\gls{ac}}}$ + \gls{mfa}$_{\mathrm{\gls{lc}}}$).}
\label{tab:air}
\end{table*}

Table \ref{tab:air} shows the results when \glspl{fu} have been placed
in isolated areas, and liquid channels have also been optimized to
decrease the temperature in the stack. With this approach, using
\gls{mfa}$_{\mathrm{\gls{lc}}}$, we can save the placement of 12
liquid channels, but still obtaining the same thermal results that
were achieved with 32. These results are shown in Figure
\ref{fig:48cores_air_20channel} and Tables
\ref{tab:optimized_channels} and \ref{tab:air}. These results have a
very important impact in terms of technological costs and fabrication
issues, and also a cost in the pumping energy required to drive liquid
through the channels. Since some \glspl{fu} are separated by the air
channels, there is a small overhead in the wirelength. This overhead
is compensated by the thermal benefits and the saving in the number of
liquid channels and, according to \cite{Changgu2011}, 5W of operating
energy for the pumping system.

The reader could miss a comparison with other techniques. We certainly
tried to compare our implementations of
\gls{mfa}$_{\mathrm{\gls{fu}*}}$ + \gls{mfa}$_{\mathrm{\gls{tsv}}}$,
\gls{mfa}$_{\mathrm{\gls{lc}}}$ and \gls{mfa}$_{\mathrm{\gls{ac}}}$
with other state-of-the-art floorplanning algorithms. However, the
most well known implementations in the literature using
representations like Combined Bucket and 2D Array (CBA), Double-Tree
and Sequence (DTS), Sequence Pair (SP) or Generalized Polish
Expression (GPE) do not perform the placement of \glspl{tsv},
\glspl{lc} or \glspl{ac}, unless \glspl{tsv} and \glspl{lc} are
explicitly included as new passive elements in the algorithm. This
would require a significant modification in
\gls{mfa}$_{\mathrm{\gls{fu}}}$, avoiding the use of
\gls{mfa}$_{\mathrm{\gls{tsv}}}$ and
\gls{mfa}$_{\mathrm{\gls{lc}}}$. In this regard, a comparative study
between CBA, DTS, SP, GPE and \gls{mfa}$_{\mathrm{\gls{fu}}}$ has been
recently published in \cite{Cuesta2015}, obviously avoiding the
placement of \glspl{tsv}, \glspl{lc} or \glspl{ac}. Because of the
different nature of CBA, DTS and SP, which are single-objective
algorithms and GPE and \gls{mfa}$_{\mathrm{\gls{fu}}}$, which are
multi-objective algorithms, two sets of experiments were performed in
\cite{Cuesta2015}. In the first set of experiments, the multi-objective
function of GPE and \gls{mfa}$_{\mathrm{\gls{fu}}}$ was transformed
into three separate optimizations minimizing temperature, wire length
and a weighted sum of both objectives, respectively. Results showed
that starting in the same conditions, i.e., starting with random
initial values, only CBA and \gls{mfa}$_{\mathrm{\gls{fu}}}$ reached
feasible solutions. In all these cases \gls{mfa}$_{\mathrm{\gls{fu}}}$
obtained the best range of temperatures and wire length. Only when
CBA, DTS, SP and GPE started with initial feasible solutions
outperformed the wire length obtained by
\gls{mfa}$_{\mathrm{\gls{fu}}}$. In the second set of experiments, in
a multi-objective optimization, GPE was not able to find a set of
feasible solutions, whereas \gls{mfa}$_{\mathrm{\gls{fu}}}$ found a
non-dominated set in all the experiments. Comparing these results with
an execution of GPE with a set of initial feasible solutions,
\gls{mfa}$_{\mathrm{\gls{fu}}}$ reached excellent temperatures but
sacrificing wire length.

\section{Conclusions} \label{sec:conclusions}
This work proposes a novel and an effective optimization process
that combines three different evolutionary algorithms to enhance our
previous and related work in the field. In this sense, our algorithms
are capable of optimizing the placement of functional units and
\glspl{tsv}, taking into account their thermal contribution to the
entire 3D stack. The optimization in the placement reduces thermal
metrics in 3D \gls{ic}. The complete floorplanner is demonstrated to
achieve excellent results when placing \glspl{fu} and
\glspl{tsv}. These results are widely improved by adding other
design solutions such as the use of active cooling, using liquid
channels whose locations have been also optimized with our proposed
evolutionary algorithms. The floorplanner interfaces with an accurate
thermal model, which calculates the results in the minimization of
thermal and reliability parameters.

Experimental results have been obtained for a realistic many-core
single-chip, resembling Niagara floorplan.

The addition of air channels to isolate thermal domains in the chip is
an important contribution presented in this paper. The isolation helps
on the cooling process, because our liquid channels can be placed in
those areas with higher temperatures. This strategy reduces the number
of deployed liquid microchannels to achieve the same thermal
profile. This reduction directly impacts on fabrication costs and on
the energy designated for the pumping system. In other words, the
creation of hot regions using air channels inside the chip has
resulted in an improvement of the thermal parameters, saving
technological and energy costs.

\section*{Acknowledgment}
This work has been partly funded by the Spanish Ministry of Economy
and Competitivity under research grant TEC2012-33892.

\bibliographystyle{elsarticle-num}
\bibliography{Bibliography}

\begin{thebibliography}{10}
\expandafter\ifx\csname url\endcsname\relax
  \def\url#1{\texttt{#1}}\fi
\expandafter\ifx\csname urlprefix\endcsname\relax\def\urlprefix{URL }\fi
\expandafter\ifx\csname href\endcsname\relax
  \def\href#1#2{#2} \def\path#1{#1}\fi

\bibitem{Choi2002}
K.~Choi, K.~Dantu, W.-C. Cheng, M.~Pedram,
  \href{http://doi.acm.org/10.1145/774572.774680}{Frame-based dynamic voltage
  and frequency scaling for a {MPEG} decoder}, in: Proceedings of the 2002
  IEEE/ACM international conference on Computer-aided design, ICCAD '02, ACM,
  New York, NY, USA, 2002, pp. 732--737.
\newblock \href {https://doi.org/10.1145/774572.774680}
  {\path{doi:10.1145/774572.774680}}.
\newline\urlprefix\url{http://doi.acm.org/10.1145/774572.774680}

\bibitem{Ogras2007}
U.~Ogras, R.~Marculescu, P.~Choudhary, D.~Marculescu, Voltage-frequency island
  partitioning for {GALS}-based networks-on-chip, in: Design Automation
  Conference, 2007. DAC '07. 44th ACM/IEEE, 2007, pp. 110 --115.

\bibitem{Cuesta2010}
D.~Cuesta, J.~L. Ayala, J.~I. Hidalgo, D.~Atienza, A.~Acquaviva, E.~Macii,
  Adaptive task migration policies for thermal control in {MPSoCs}, in: ISVLSI,
  2010, pp. 110--115.

\bibitem{Shang2004}
L.~Shang, L.-S. Peh, A.~Kumar, N.~K. Jha, Thermal modeling, characterization
  and management of on-chip networks, in: Proceedings of the 37th annual
  IEEE/ACM International Symposium on Microarchitecture, MICRO 37, IEEE
  Computer Society, Washington, DC, USA, 2004, pp. 67--78.

\bibitem{Jonggook:TemperatureGradient:99}
K.~Jonggook, V.~Tyree, C.~Crowell, Temperature gradient effects in
  electromigration using an extended transition probability model and
  temperature gradient free tests. i. transition probability model, in:
  Integrated Reliability Workshop Final Report, 1999. IEEE International, 1999,
  pp. 24--40.

\bibitem{Cong2004}
J.~Cong, J.~Wei, Y.~Zhang, A thermal-driven floorplanning algorithm for {3D}
  {ICs}, in: Proceedings of the 2004 IEEE/ACM International conference on
  Computer-aided design, ICCAD '04, IEEE Computer Society, Washington, DC, USA,
  2004, pp. 306--313.

\bibitem{Healy2007}
M.~Healy, M.~Vittes, M.~Ekpanyapong, C.~S. Ballapuram, S.~K. Lim, H.-H.~S. Lee,
  G.~H. Loh, Multiobjective microarchitectural floorplanning for {2-D} and
  {3-D} {ICs}, Computer-Aided Design of Integrated Circuits and Systems, IEEE
  Transactions on 26~(1) (2007) 38--52.
\newblock \href {https://doi.org/10.1109/TCAD.2006.883925}
  {\path{doi:10.1109/TCAD.2006.883925}}.

\bibitem{Wong2006}
E.~Wong, S.~Lim, {3D} floorplanning with thermal vias, in: DATE, 2006, pp.
  878--883.

\bibitem{DelValle2010}
P.~G.~D. Valle, D.~Atienza, Emulation-based transient thermal modeling of
  {2D/3D} systems-on-chip with active cooling, Microelectronics Journal 1
  (2010) 564--571.

\bibitem{Coskun2009}
A.~Coskun, J.~Ayala, D.~Atienza, T.~Rosing, Y.~Leblicini, Dynamic thermal
  management in {3D} multicore architectures, Design, Automation and Test in
  Europe (2009) 1410--1415.

\bibitem{Sabry2011}
M.~Sabry, A.~Sridhar, D.~Atienza, Y.~Temiz, Y.~Leblebici, S.~Szczukiewicz,
  N.~Borhani, J.~Thome, T.~Brunschwiler, B.~Michel, Towards thermally-aware
  design of {3D} {MPSoCs} with inter-tier cooling, in: Design, Automation Test
  in Europe Conference Exhibition (DATE), 2011, 2011, pp. 1 --6.

\bibitem{Berntsson2004}
J.~Berntsson, M.~Tang, A slicing structure representation for the multi-layer
  floorplan layout problem, in: EvoWorkshops, 2004, pp. 188--197.

\bibitem{Tang2007}
M.~Tang, X.~Yao, A memetic algorithm for {VLSI} floorplanning, IEEE
  Transactions on Systems, Man, and Cybernetics, Part B 37~(1) (2007) 62--69.

\bibitem{Cuesta2013}
D.~Cuesta, J.~L. Risco-Martín, J.~L. Ayala, J.~I. Hidalgo, 3d thermal-aware
  floorplanner using a {MOEA} approximation, Integration, the VLSI Journal
  46~(1) (2013) 10--21.
\newblock \href {https://doi.org/10.1016/j.vlsi.2012.04.003}
  {\path{doi:10.1016/j.vlsi.2012.04.003}}.

\bibitem{Brooks2007}
D.~Brooks, R.~P. Dick, R.~Joseph, L.~Shang,
  \href{http://ieeexplore.ieee.org/lpdocs/epic03/wrapper.htm?arnumber=4292056}{{Power,
  Thermal, and Reliability Modeling in Nanometer-Scale Microprocessors}}, IEEE
  Micro 27~(3) (2007) 49--62.
\newblock \href {https://doi.org/10.1109/MM.2007.58}
  {\path{doi:10.1109/MM.2007.58}}.
\newline\urlprefix\url{http://ieeexplore.ieee.org/lpdocs/epic03/wrapper.htm?arnumber=4292056}

\bibitem{Ayala2012}
J.~L. Ayala, A.~Sridhar, D.~Atienza, Y.~Leblebici, Design {T}echnology for
  {H}eterogeneous {E}mbedded {S}ystems, Vol.~1 of Lecture Notes in Computer
  Science, Springer, 2012, Ch. Through {S}ilicon {V}ia-{B}ased {G}rid for
  {T}hermal {C}ontrol in 3{D} {C}hips, pp. 303--320.

\bibitem{Sridar2010}
A.~Sridar, A.~Vicenzi, M.~Ruggiero, T.~Brunschwiler, D.~Atienza, {3D-ICE}: Fast
  compact transient thermal modeling for {3D} {ICs} with inter-tier liquid
  cooling, Computer-Aided Design (ICCAD) (2010) 463--470.

\bibitem{Deb2002}
K.~Deb, A.~Pratap, S.~Agarwal, T.~Meyarivan, A fast and elitist multiobjective
  genetic algorithm: {NSGA-II}, IEEE Transactions on Evolutionary Computation
  6~(2) (2002) 182--197.

\bibitem{Sivanandam2007}
S.~N. Sivanandam, S.~N. Deepa, Introduction to Genetic Algorithms, Springer,
  2007.

\bibitem{Weldezion2009}
A.~Weldezion, et~Al, Bandwidth optimization for through silicon via {(TSV)}
  bundles in {3D} integrated circuits, in: Design, Automation and Test in
  Europe, 2009.

\bibitem{Changgu2011}
C.~Lee, M.~Liamini, L.~Frechette, A silicon microturbopump for a rankine-cycle
  power-generation microsystem, Microelectromechanical Systems, Journal of
  20~(1) (2011) 326--338.

\bibitem{Niagara:Niagara}
http://www.opensparc.net/pubs/preszo/07/n2isscc.pdf.

\bibitem{Cuesta2015}
A.~Cuesta-Infante, J.~M. Colmenar, Z.~Bankovic, J.~L. Risco-Martín, M.~Zapater,
  J.~I. Hidalgo, J.~L. Ayala, J.~M. Moya,
  \href{http://www.sciencedirect.com/science/article/pii/S0925231214012429}{Comparative
  study of meta-heuristic 3{D} floorplanning algorithms}, Neurocomputing 150,
  Part A~(0) (2015) 67 -- 81.
\newblock \href
  {https://doi.org/http://dx.doi.org/10.1016/j.neucom.2014.06.078}
  {\path{doi:http://dx.doi.org/10.1016/j.neucom.2014.06.078}}.
\newline\urlprefix\url{http://www.sciencedirect.com/science/article/pii/S0925231214012429}

\end{thebibliography}

\end{document}